\documentclass[aip,jcp,reprint,letterpaper,onecolumn]{revtex4-1}
\usepackage{amssymb}
\usepackage{amsmath}
\usepackage[usenames,dvipsnames]{xcolor}
\usepackage{graphicx}
\usepackage{float}
\usepackage{subcaption}
\usepackage[top=0.9in, bottom=0.9in,left=0.9in,right=0.9in]{geometry}
\usepackage{natbib}

\newcommand{\abs}[1]{\left|#1\right|}
\begin{document}
	\title
	{Well-behaved versus ill-behaved density functionals for single bond dissociation: Separating success from disaster functional by functional for stretched H$_2$}
	\author{Diptarka Hait}
	%
	\thanks{These authors contributed equally to this work.}
	\affiliation
	{{Kenneth S. Pitzer Center for Theoretical Chemistry, Department of Chemistry, University of California, Berkeley, California 94720, USA}}
	\author{Adam Rettig}
	\thanks{These authors contributed equally to this work.}
	\affiliation
	{{Kenneth S. Pitzer Center for Theoretical Chemistry, Department of Chemistry, University of California, Berkeley, California 94720, USA}}
	\author{Martin Head-Gordon}
	\email{mhg@cchem.berkeley.edu}
	\affiliation
	{{Kenneth S. Pitzer Center for Theoretical Chemistry, Department of Chemistry, University of California, Berkeley, California 94720, USA}}
	\affiliation{Chemical Sciences Division, Lawrence Berkeley National Laboratory, Berkeley, California 94720, USA}
   
\begin{abstract}
	Unrestricted DFT methods are typically expected to describe the homolytic dissociation of nonpolar single bonds in neutral species with qualitative accuracy, due to the lack of significant delocalization error. We however find that many widely used density functional approximations fail to describe features along the dissociation curve of the simple H$_2$ molecule. This is not an universal failure of DFT in the sense that many classic functionals like PBE and B3LYP give very reasonable results, as do some more modern methods like MS2. However, some other widely used functionals like B97-D (empirically fitted) and TPSS (non-empirically constrained) predict qualitatively wrong static polarizabilities, force constants and some even introduce an artificial barrier against association of independent H atoms to form H$_2$. The polarizability and force constant prediction failures appear to stem from incomplete spin localization into individual H atoms beyond the Coulson-Fisher point, resulting in `fractionally bonded' species where the ionic contributions to the Slater determinant are not completely eliminated, unlike the case of unrestricted Hartree-Fock. These errors therefore appear to be a consequence of poor self-consistent density prediction by the problematic functional. The same reasons could potentially lead to spurious barriers towards H atom association, indirectly also leading to incorrect forces. These unphysicalities suggest that the use of problematic functionals is probably unwise in \textit{ab initio} dynamics calculations, especially if strong electrostatic interactions are possible. 
\end{abstract}
	\maketitle
    
	\section{Introduction}
	Kohn-Sham density functional theory (KS-DFT)\cite{kohn1965self} is the most widely used electronic structure method used to calculate energies and properties of molecules and extended materials\cite{becke2014perspective,jones2015density,mardirossian2017thirty}. KS-DFT \textit{assumes} that the exact electron density of a given system can be reproduced by a single Slater determinant (or, in the case of double hybrid functionals, a single Slater determinant with small corrections) and is consequently quite successful for many systems where one Slater determinant heavily dominates the wave function (i.e. single-reference systems). This is not to say that KS-DFT is perfect for all single-reference systems---indeed, delocalization error\cite{perdew1982density,perdew1985kohn,zhang1998challenge} leads to catastrophic failures in many cases \cite{vydrov2007tests,cohen2008fractional,mori2008localization,dreuw2005single,dreuw2003long}. Nonetheless, these errors are well characterized and can often be mitigated by careful choice of functionals\cite{hait2018delocalization}.
	
	The performance of KS-DFT is considerably more questionable for multi-reference (MR) systems where many Slater determinants make substantial contributions to the exact wave function, due to large differences between the true kinetic energy and the non-interacting approximation to it that KS-DFT obtains from a single determinant. However, this also does not imply that KS-DFT cannot be applied to any MR system, especially when spin-symmetry broken, unrestricted determinants are employed. The case of stretched single bonds offer the simplest example: bonded electrons would try to localize on the individual fragments in the highly stretched limit, and a spin-restricted determinant would qualitatively fail to reproduce the density as it would force up and down spins into the same spatial orbitals. However, an unrestricted formalism would permit up and down spins to localize in different spatial locations, allowing a physically correct (albeit spin-polarized) description of the density to arise. Unrestricted single determinant methods consequently break spin symmetry when bonds are stretched beyond a limit called the Coulson-Fischer (CF) point\cite{coulson1949xxxiv}. The determinant however ceases to be an eigenstate of $\hat{S}^2$, as the spin polarized solution would be made out of approximately equal amounts of singlet and triplet (plus smaller intrusions from higher spin states, where possible). This appears to be problematic on the surface, since the true eigenstate for the non-relativistic Hamiltonian (that KS-DFT is attempting to solve for) should be spin pure. However, the spins on the two fragments will be fully noninteracting in the broken bond limit, and the singlet and triplet states of the noninteracting fragments should be equivalent, making this spin polarized state adequate for prediction of energetics and properties not directly related to spins. Indeed, this is the protocol successfully used to evaluate atomization energies, since it is size-consistent (in the sense that the energy $E_{AB}$ of a system with a highly stretched $A$---$B$ single bond will asymptote to the sum of energies $E_A$ and $E_B$ of isolated individual fragments $A$ and $B$ respectively). We must however note that efforts have been made to develop  `strong correlation' functionals that operate on spin-restricted densities for treating MR problems without spin contamination issues\cite{becke2013density,johnson2013density,kong2015density,laqua2018communication}. Unfortunately, such methods are typically not size-consistent\cite{becke2013density,johnson2013density,kong2015density,laqua2018communication} although they do succeed in reducing the magnitude of the size-consistency energy error substantially relative to traditional spin-restricted methods. 
	
	The intermediate regime where the spins on the fragments are not quite independent has more potential to be problematic for spin unrestricted methods. However in practice, single reference unrestricted Hartree-Fock (UHF) theory gives qualitatively acceptable results in this regime\cite{szabo2012modern}, showing that single reference theories can generate reasonable values even for such fairly MR problems, by taking advantage of the weak interaction between the polarized spins. Unrestricted KS-DFT (UKS) methods, on the other hand, are known to make much more problematic predictions for stretched single bonds, but generally on account of well understood delocalization effects. It is for instance quite well known that the dissociation curves for X$_2^+$ (where X is any monovalent group like H or CH$_3$) species will have a barrier for fragment association, and in extreme cases have a \textit{negative} dissociation energy \cite{grafenstein2004impact} due to overstabilization of fractional charges in the dissociation limit. Similarly, local functionals cause polar bonds to dissociate into fractionally charged constituents\cite{Dutoi2006,Ruzsinszky2006} and even hybrid functionals delocalize charge too much, resulting in a dipole moment that decays too slowly as the bond is stretched\cite{hait2018accurate}.  
	
	It is however typically assumed that such delocalization driven unphysical behavior is absent from dissociation curves of nonpolar single bonds in neutral molecules. This is a quite reasonable line of thought as there is no intrinsic driving force towards fractional charges (unlike the case of charged species or polar bonds). It is indeed possible that the bond would dissociate to fragments with fractional spins\cite{cohen2008fractionalspin} instead of fractional charges, but this is also rather unlikely as single reference HF and KS-DFT methods tend to overestimate the energy of species with fractional spins\cite{cohen2008fractionalspin}, biasing the calculation against any such possibility in the dissociation limit. Both UHF and UKS theory thus are commonly expected to dissociate nonpolar single bonds in neutral molecules in a qualitatively correct manner, and generate the right dissociation limit of neutral fragments with half-integer spin. This permits widespread use of KS-DFT in studying reactive trajectories, as a reasonable description of bond formation and dissociation processes are expected. 
    
	 A qualitatively correct description of the stretched bond regime should also ensure reasonable prediction of molecular properties. Indeed, properties like static polarizabilities and force constants should be useful for revealing complications in the description of electronic structure. These properties are second derivatives of the molecular energy with respect to parameters like external electric fields or bond length and should consequently magnify any issues present in the underlying energy/density predictions, that would not be apparent by just looking at the potential energy surface. Such second derivative properties are therefore useful probes of high sensitivity for assessing the quality of electronic structure methods for stretched nonpolar bonds, similar to how the first derivative property dipole moment helps reveal catastrophic density failures for stretched polar bonds\cite{Kurlancheek2009,hait2018accurate}. 
     
     	        	\begin{figure}[htb!]
        			\includegraphics[width=0.5\linewidth]{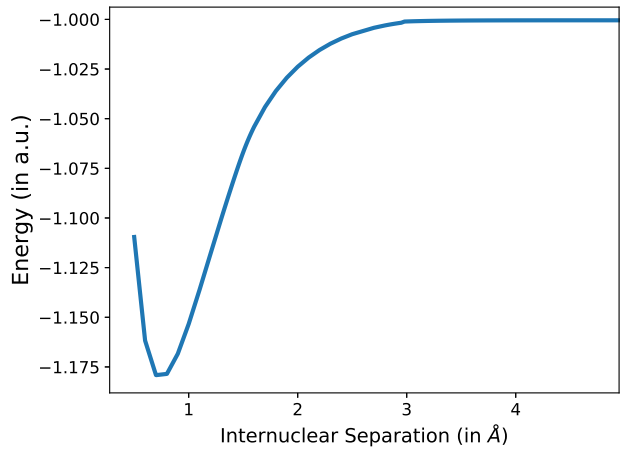}
        		\caption{Energy prediction for stretched H$_2$ by TPSS.}
        		\label{fig:tpssE}
        	\end{figure}
     
     Our study of second derivative properties for stretched H$_2$ reveals that while HF (or more specifically, UHF) and several density functionals like PBE\cite{PBE} yield reasonable behavior for stretched H$_2$, many widely used approximations yield unphysical static polarizabilities and force constants, with several also predicting an unphysical barrier for association of two H atoms to form H$_2$. The underlying potential energy surfaces look quite smooth overall and do not immediately reveal any significant unphysicalities aside from potentially a barrier. Even such barriers are typically $<2.5$ \% of the bond dissociation energy, and therefore quite hard to identify from  a first glance, if it is even present. TPSS\cite{tpss} for instance has been long been claimed to give reasonable energy predictions for stretched H$_2$\cite{ruzsinszky2005binding}, and Fig \ref{fig:tpssE} shows it to be quite barrier free. The static polarizability and force constants however reveal significant complications between the CF point and the dissociation limit (as is discussed later), indicating that exploration of second derivative properties was essential to identify the points of failure. 
     
     In the next section, we detail the computational methods employed by us to reach our conclusions. We subsequently discuss what we believe constitutes reasonable behavior for a KS-DFT method for stretched single bonds, and show that  problematic features in static polarizability and force constants are predicted by both empirically fitted functionals like B97-D\cite{b97d} and M06-L\cite{m06l}, as well as nonempirical ones like TPSS\cite{tpss}, indicating that overparameterization alone is only one potential origin of the issue. We then highlight that these second derivative property prediction failures are connected to unphysicalities in the smallest eigenvalue of the Hessian of the energy with respect to orbital rotations, and incomplete spin localization beyond the CF point. The latter suggests that a fractional bond is preserved even at large internuclear separation, adding a substantial `ionic' contribution to the UKS wave function, instead of the expected `covalent' behavior. Afterwards, we show that use of `physically correct' (i.e. fully spin polarized) densities could mitigate this problem, and conclude by discussing the barriers towards H atom association for H$_2$ formation that many functionals predict, suggesting that the problematic functionals ought not to be used to study the \textit{ab initio} dynamics involving bond formation or rupture, especially when better behaved alternatives are available.

	\section{Computational Methods}
	All calculations were done with the Q-Chem\cite{QCHEM4} software package. HF/DFT calculations used spin-unrestricted orbitals, the quintuiple zeta aug-pc-4 basis\cite{jensen2001polarization,jensen2002polarization,jensen2002polarizationiii} and an integration grid using 250 radial and 974 Lebedev angular points for the local exchange-correlation integrals, unless specified otherwise. We note that use of smaller grids or basis sets yielded essentially the same behavior, indicating that the unphysicalities are unlikely to stem from basis set or grid incompleteness errors. Stability analysis was performed for every solution to ensure that the Slater determinant was a (local) minimum with respect to occupied-virtual rotations. The CISD calculations were performed with the aug-cc-pVTZ\cite{dunning1989gaussian} basis in order to get a qualitative understanding of exact behavior. Static polarizabilities were calculated from a central finite difference formula using an applied electric field strength of $10^{-3}$ a.u. Exploratory tests with  different field strengths yielded the same behavior, indicating negligible finite difference step size errors.
	\section{Results and Discussions}
	\subsection{Qualitatively correct functionals}
	\begin{figure}[htb!]
		\begin{minipage}{0.48\textwidth}
			\includegraphics[width=\linewidth]{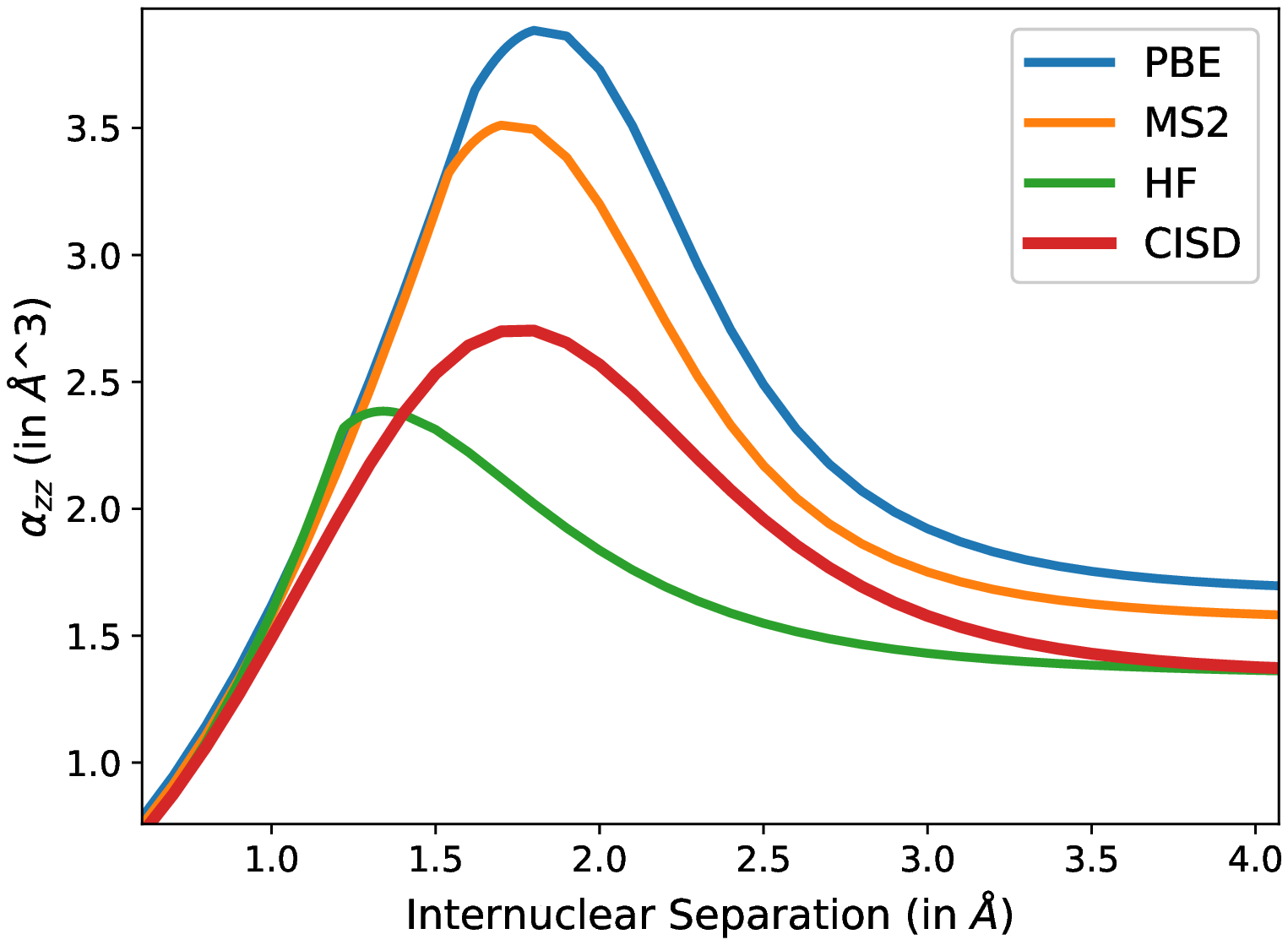}
		\end{minipage}
			\begin{minipage}{0.48\textwidth}
				\includegraphics[width=\linewidth]{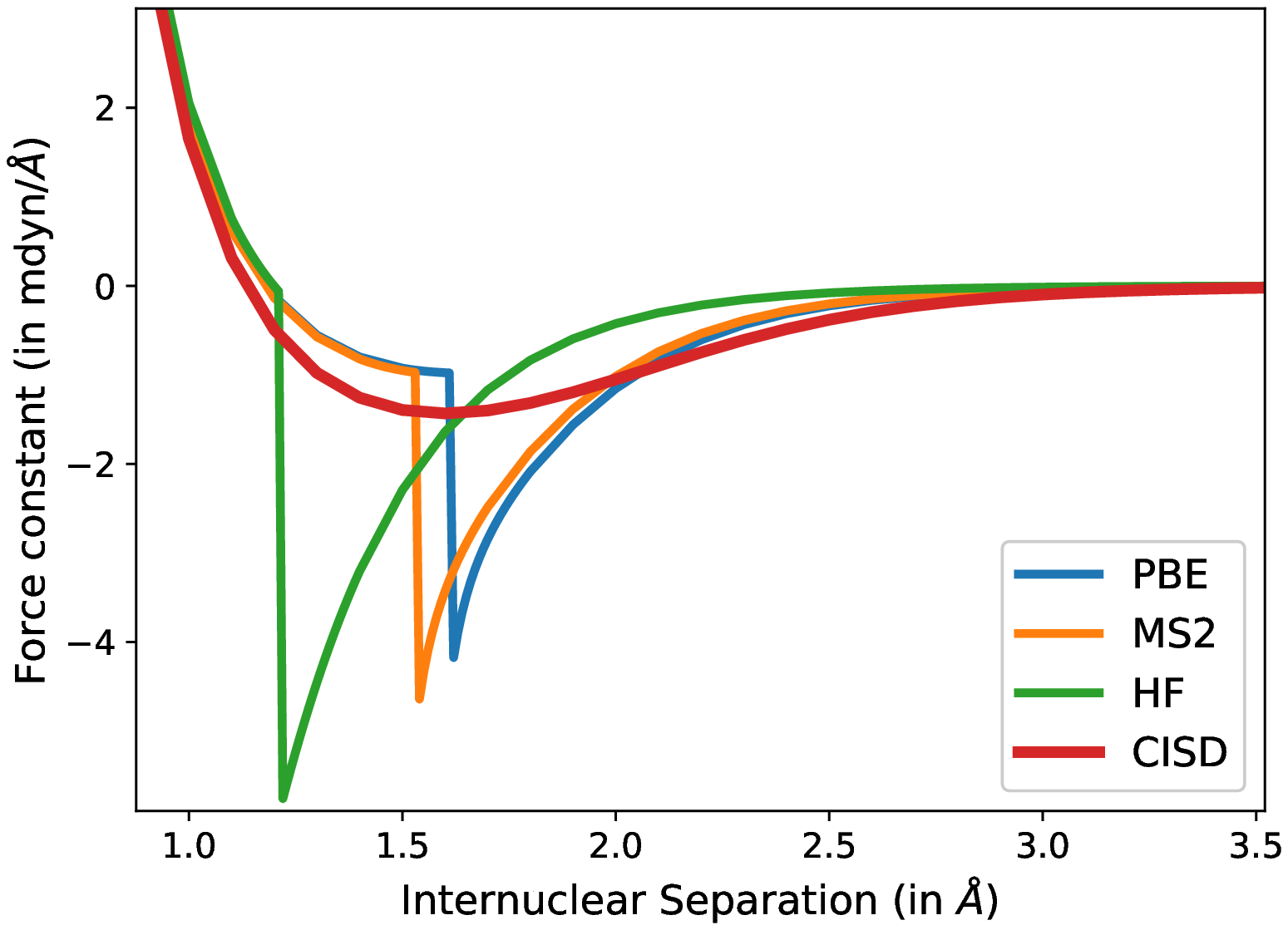}
			\end{minipage}
		\caption{Static polarizability along bond axis (left) and force constants for bond stretching (right) predicted by three `qualitatively acceptable' methods (HF, PBE and MS2), compared to exact (CISD) values for stretched H$_2$. The polarizability predictions by all three single determinant methods are qualitatively fine, save a barely perceptible kink at the CF point. There is a discontinuity in force constant predictions at the CF point for all three but the behavior is reasonable otherwise. }
		\label{fig:polaracc}
	\end{figure}
    We consider a functional to be qualitatively correct if it yields the same general behavior as exact quantum mechanics at all points in the dissociation curve, except possibly at the CF point. Single reference methods can predict non-analytic behavior at the CF point in general\cite{szabo2012modern,Kurlancheek2009,hait2018accurate}, and therefore we take HF as the baseline for what constitutes `reasonable' in the immediate neighborhood of the  CF point. This choice stems from HF giving qualitatively right behavior at all other points, and because it seems reasonable to expect UKS to at least fare no worse than the best possible single determinant wave function method (i.e. HF). Indeed, more advanced wave function methods like orbital optimized second order M{\o}ller-Plesset perturbation theory (OO-MP2)\cite{lochan2007orbital,neese2009assessment} have to be explicitly regularized to recover HF like behavior at the CF point\cite{stuck2013regularized,lee2018regularized}. An UKS method that yields behavior closer to the exact results than UHF around the CF point should also be viewed as qualitatively accurate in that neighborhood (although failure elsewhere would reflect poorly on the method in general). 

Our focus on static polarizabilities and force constants indicate a need to understand the mathematical behavior of properties that can be expressed as second derivatives of the energy. Specifically, given a wave function with orbital degrees of freedom $\theta$, we can say that the rate of change of an observable $A$ against some parameter $x$ can be expressed as:
\begin{align}
\dfrac{dA}{dx}=\dfrac{\partial A}{\partial x}+\left(\dfrac{\partial A}{\partial \theta}\right)_x\dfrac{\partial \theta}{\partial x}
\end{align}
from the chain rule, where $\dfrac{\partial \theta}{\partial x}$ represents the rate of change of orbitals. For a variationally optimized energy $E$, we therefore have:
\begin{align}
\dfrac{dE}{dx}=\dfrac{\partial E}{\partial x}
\end{align} 
as the response of the energy to orbital rotation $\left(\dfrac{\partial E}{\partial \theta}\right)_x=0$ from the Hellman-Feynman theorem. However, this does not apply to energy second derivatives, which can be expressed as:
\begin{align}
\dfrac{d^2E}{dxdy}=\dfrac{\partial^2 E}{\partial x\partial y}+\dfrac{\partial^2 E}{\partial x\partial \theta}\dfrac{\partial \theta}{\partial y}=\dfrac{\partial^2 E}{\partial x\partial y}+\dfrac{\partial^2 E}{\partial x\partial \theta}\left[\dfrac{\partial^2 E}{\partial \theta^2}\right]^{-1}\dfrac{\partial^2 E}{\partial y\partial \theta}\label{eqnpol}
\end{align}
where $\dfrac{\partial^2 E}{\partial \theta^2}$ is the Hessian of the energy with respect to orbital rotations. This Hessian however has a zero eigenvalue at the CF point of typical single determinant methods\cite{szabo2012modern}, due to barrierless degeneracy of the spin-unpolarized restricted and spin polarized solutions. Consequently, second derivative properties may be undefined at the CF point itself due to the inversion of a singular matrix, and is discontinuous in the neighborhood due to discontinuity in $\dfrac{\partial \theta}{\partial x}$ (the rate of change of orbital degrees of freedom with respect to the parameter $x$) on both sides of the CF point. This is indeed observed for the force constant for UHF, as can be seen on the right panel of Fig \ref{fig:polaracc}. 
\newline \indent
However, the left panel of Fig \ref{fig:polaracc} shows that the polarizability (where the applied electric field $\mathcal{E}=x=y$) does not appear to show such a discontinuity at the CF point for H$_2$, even though a derivative discontinuity (kink) is present. This is a consequence of the lack of a permanent dipole moment for the molecule. The spin polarization transition therefore does not affect the polarity, ensuring that the rate of change of the dipole $\dfrac{\partial \mu }{\partial \theta}=\dfrac{\partial^2 E }{\partial \theta\partial \mathcal{E}}=0$ along the eigenvector of the Hessian leading to the transition. This eigenvector however is associated with the zero eigenvalue, and therefore no contributions from the singular term survives in Eqn \ref{eqnpol}, leaving behind a continuous function. However, polar bonds can have nonzero $\dfrac{\partial^2 E }{\partial \theta\partial \mathcal{E}}$ along the spin-polarization transition eigenvector and so a discontinuous polarizability can be observed for such systems (and indeed, we observed it for hydrogen fluoride with UHF).

    In summary, a derivative discontinuity (kink) at the CF point is acceptable for polarizabilities of nonpolar H$_2$, while a discontinuity at the CF point is permissible for the force constant. We find that the PBE GGA functional and the MS2\cite{ms2} meta-GGA (mGGA) functional satisfy the above criteria for the properties we examine (as can be seen from Fig \ref{fig:polaracc}), and are therefore taken as a baseline for acceptable behavior for functionals from the second and third rungs of Jacob's ladder\cite{perdew2001jacob} respectively. We do however note that these two functionals systematically overestimate polarizabilities on account of delocalization error\cite{hait2018delocalization,hait2018accuratepolar} while HF tends to underestimate polarizabilities in the stretched bond regime due to a systematic bias towards electron localization stemming from missing correlation energy. The hybrid variants PBE0\cite{pbe0} and MS2h\cite{ms2} also yield sensible results, as does the the popular B3LYP\cite{b3lyp}.
    
    \subsection{Static Polarizabilities}
    The willingness of the electron density in the bonded region to `polarize' is a function of the strength of the chemical bond. A strong bond would be unwilling to distort the natural density and thus have low polarizability. Electrons in a stretched bond are much more likely to be polarizable, as they are weakly held by the nuclei and can be displaced in whichever direction an electric field nudges them to. Beyond a certain point however, the bonded pair localizes into individual fragments and the polarizability decays to the fragment limit.   This behavior is seen for instance in Fig \ref{fig:polaracc} (left panel), where the polarizability of H$_2$ rises with the weakening of the bond via stretching, until the electrons start to localize on individual H atoms, leading to an asymptotic decline to the atomic limit. Reasonable functionals like PBE are expected to give similar results as well, with a kink at the CF point heralding the start of density localization. Unfortunately, we discover that many functionals that are widely used in quantum chemistry yield an inaccurate description of static polarizability for stretched bonds, signaling inaccurate description of the bonding process. 
    	\begin{figure}[htb!]
    		\begin{minipage}{0.48\textwidth}
    			\includegraphics[width=\linewidth]{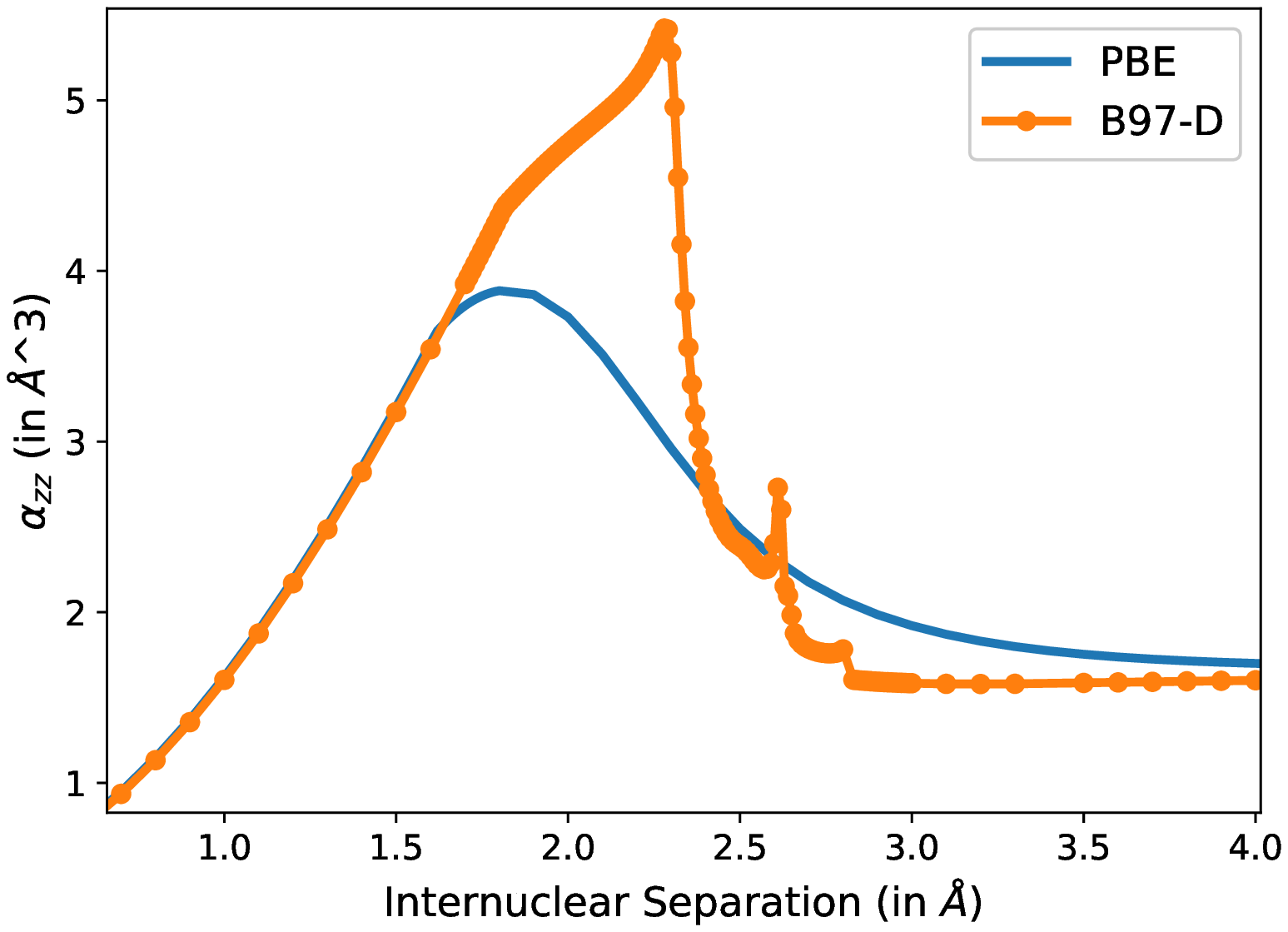}
    		\end{minipage}
    		\begin{minipage}{0.48\textwidth}
    			\includegraphics[width=\linewidth]{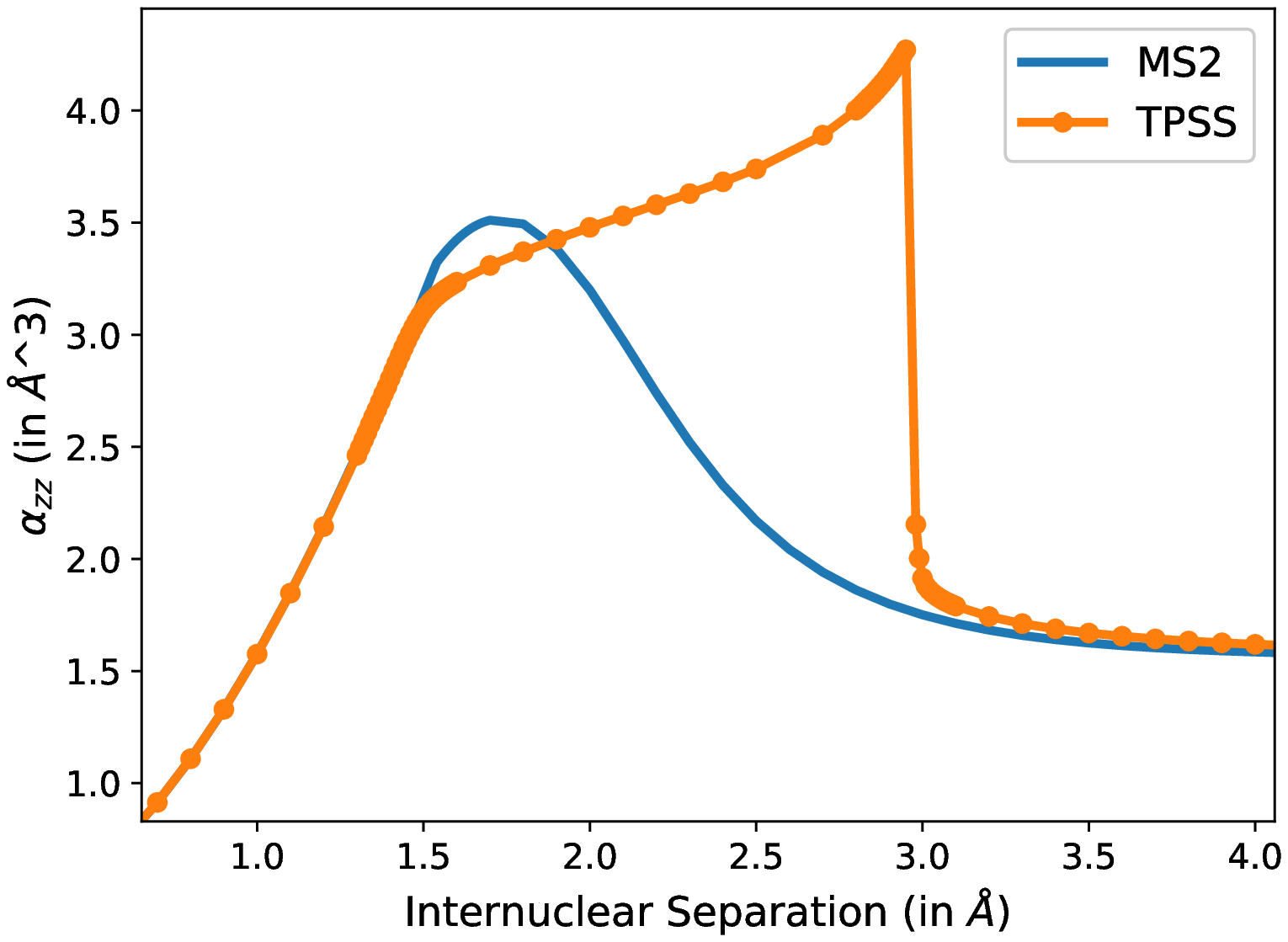}
    		\end{minipage}
    		\caption{Static polarizability along bond axis predicted by B97-D (left) and TPSS (right), with an acceptable functional from the same rung of Jacob's ladder as a reference. Markers for points plotted are given to highlight that the problematic regions were heavily sampled, and so the features are not interpolation/plotting artifacts. The unphysical features here are not local to the CF point region (1.8 {\AA} for B97-D and 1.4 {\AA} for TPSS), although their onset roughly corresponds to that neighborhood. }
    		\label{fig:polarinacc}
    	\end{figure}
    	
    The behavior of the B97-D GGA and the TPSS mGGA are especially troubling, as they highlight two highly similar and yet distinct modes of catastrophic failure. B97-D is an empirically parameterized functional with 9 parameters, which is regarded as one of the best GGAs for prediction of energetics\cite{goerigk2017look,mardirossian2017thirty} and is superior to PBE for estimating dipole moments\cite{hait2018accurate} and static polarizabilities at equilibrium geometries\cite{hait2018accuratepolar} . Nonetheless,  it predicts monotonically increasing static polarizability well beyond the CF point at 1.8 {\AA} (as can be seen on the left panel of Fig \ref{fig:polarinacc}), indicating that the bonded density was not cleanly localizing to separate fragments. This is followed by a very rapid (though not discontinuous) drop to near the atomic asymptote around 2.3 {\AA}, a second, less prominent peak near 2.6 {\AA} and a small discontinuity at 2.8 {\AA}, suggesting that the UKS density was not localizing into the separate H atoms in a clean, UHF like manner. 
   
    TPSS has a similar monotonic increase well beyond the CF point at 1.4 {\AA}, which lasts till about $r=3$ {\AA}, beyond which it drops discontinuously to a value very close to the atomic asymptote, and afterwards decays smoothly. This transition appears to be discontinuous (as opposed to merely very rapid, which was the case for B97-D) based on sampling at intervals of 0.01 {\AA}, which is quite troubling as it suggests a dramatic, discontinuous change in the underlying density. At any rate, TPSS seems ill suited for finding electric field responses for H$_2$ at stretched geometries.

        	\begin{figure}[htb!]
        			\includegraphics[width=0.5\linewidth]{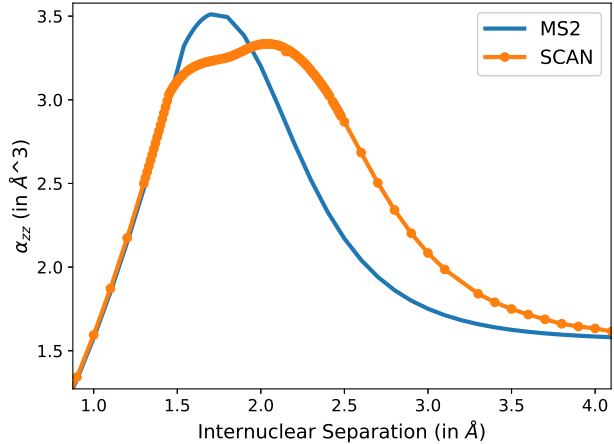}
        		\caption{Static polarizability along bond axis predicted by SCAN (right), alongside reference MS2 values.}
        		\label{fig:polarscan}
        	\end{figure}
        	
    The more recently developed SCAN\cite{SCAN} functional gives better performance than TPSS, but is not entirely free from faults. Fig \ref{fig:polarscan} shows that SCAN predicts two very closely spaced peaks in the polarizability curve, which is not quite what one ought to expect. This twin peak structure essentially leads to a plateau instead of a peak in the polarizability values along the dissociation curve, suggesting that the density localization is not happening as quickly as in HF, although is likely still happening in a smooth manner, unlike the case of TPSS. This strange behavior is nonetheless worthwhile to keep in mind while developing future meta-GGA functionals, as it shows that all of the strong nonempirical constraints and appropriate norms employed to develop SCAN were not by themselves sufficient to reproduce reasonable behavior for stretched H$_2$, which simpler functionals like PBE can readily achieve. It is also important to note that these unphysical features of TPSS and SCAN persist in their hybrid variants TPSSh\cite{tpssh} and SCAN0\cite{scan0} (although the polarizability values are lower in those cases), indicating that this behavior is not solely a consequence of some form of delocalization error.
 
         	\begin{figure}[htb!]
         		\includegraphics[width=0.5\linewidth]{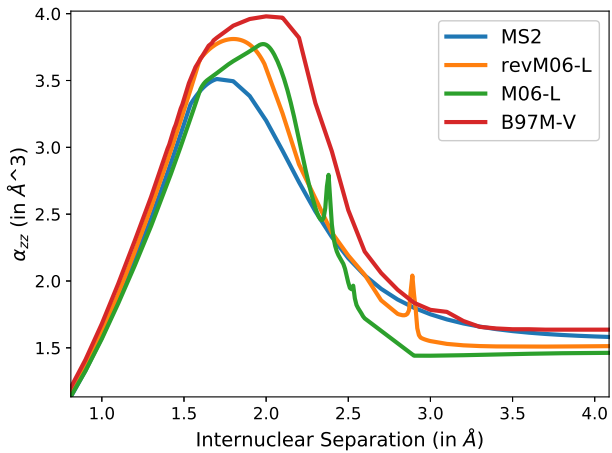}
         		\caption{Static polarizability along bond axis predicted by empirically trained functionals M06-L, revM06-L and B97M-V, alongside MS2 reference values. }
         		\label{fig:polarempirical}
         	\end{figure}
         	
      It is also interesting to consider the performance of empirical mGGAs like M06-L\cite{m06l}, revM06-L\cite{revm06l} and B97M-V\cite{b97mv} against SCAN and TPSS. These functionals give reasonable behavior for most parts of the dissociation curve, but regions of local failure are very much visible in the form of unexpected local maximums.  M06-L and revM06-L in fact have quite sharp peaks in the plot, while B97M-V has only a very small bump at around 3 {\AA} separation. 
      
\subsection{Force Constants}
The failures of these functionals in predicting static polarizabilities at stretched geometries naturally leads to the question as to whether any other second-order property (i.e. based on a second derivative of the energy) shows similar unphysical behavior. The force constant (i.e. the negative of the second derivative of energy with respect to bond stretching) is therefore a natural observable to investigate. 
    	\begin{figure}[htb!]
    		\begin{minipage}{0.48\textwidth}
    			\includegraphics[width=\linewidth]{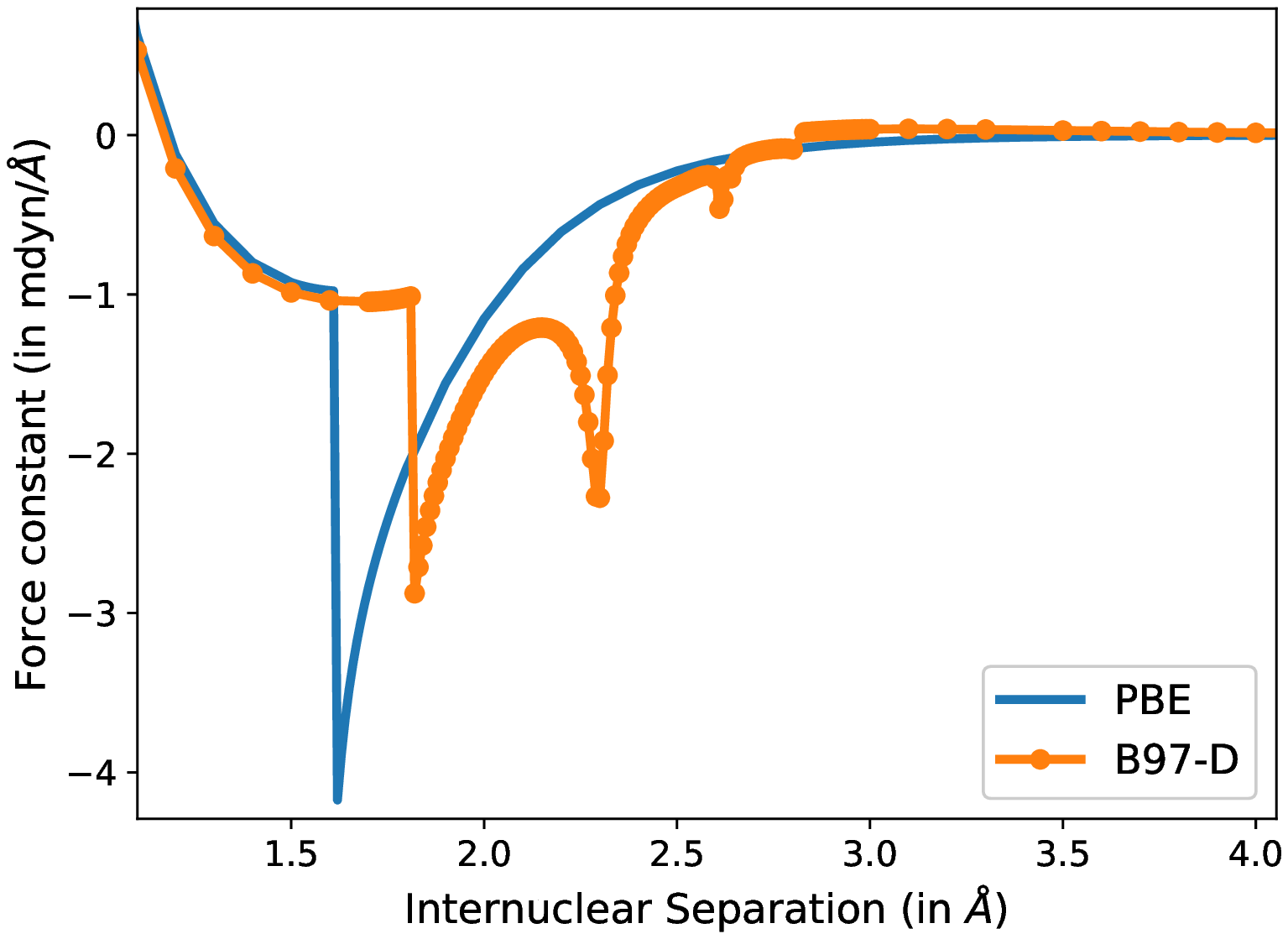}
    		\end{minipage}
    		\begin{minipage}{0.48\textwidth}
    			\includegraphics[width=\linewidth]{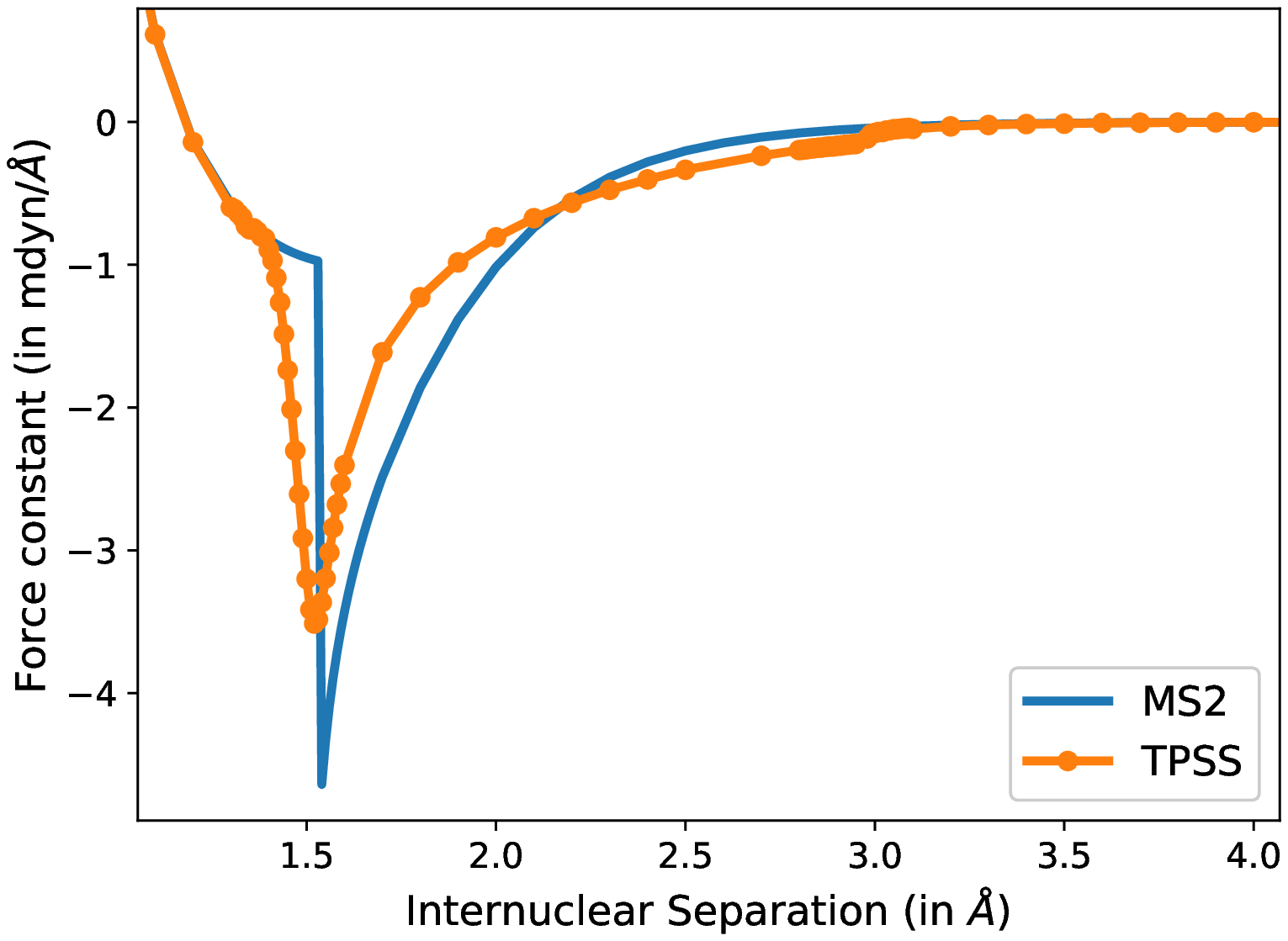}
    		\end{minipage}
    		\caption{Force constant for bond stretching predicted by B97-D (left) and TPSS (right), with an acceptable functional from the same rung of Jacob's ladder as a reference.}
    		\label{fig:fconstinacc}
    	\end{figure}

The left panel of Fig \ref{fig:fconstinacc} reveals that the B97-D force constant plot has a second peak at around 2.3 {\AA}, which is not a feature present in the exact, HF or PBE/MS2 plots, which corresponds precisely to the point where the dramatic drop of polarizability occurs in the left panel of Fig \ref{fig:polarinacc}. Furthermore, satellite structures are present further along the dissociation curve, which spatially correspond to the same internuclear separation as the secondary unphysical features in the static polarizability curve. It is therefore not unreasonable to conclude that the qualitative errors in the force constant and polarizability stem from the same factors. 

The case of TPSS (right panel of Fig \ref{fig:fconstinacc}) is more interesting in that there does not appear to be a discontinuity in force constant values at the CF point, unlike HF, PBE or MS2! A \textit{derivative} discontinuity appears to exist, but the force constant value itself continuously  changes with bond stretch. The CF point, and associated discontinuities are ultimately artifacts due to the onset of spin polarization in single determinant approximations that exact quantum mechanics does not predict, and so the absence of a normal CF point is not a sign of unphysical behavior in itself. However, the unphysical polarizability predictions shown in the right panel of Fig \ref{fig:polarinacc} and the subtle discontinuity in the force constant plot  around 3 {\AA} suggest that the lack of a normal CF point is more likely to be a symptom of a problem stemming from ineffective localization of spins than a desirable feature.

    	\begin{figure}[htb!]
    		\begin{minipage}{0.48\textwidth}
    			\includegraphics[width=\linewidth]{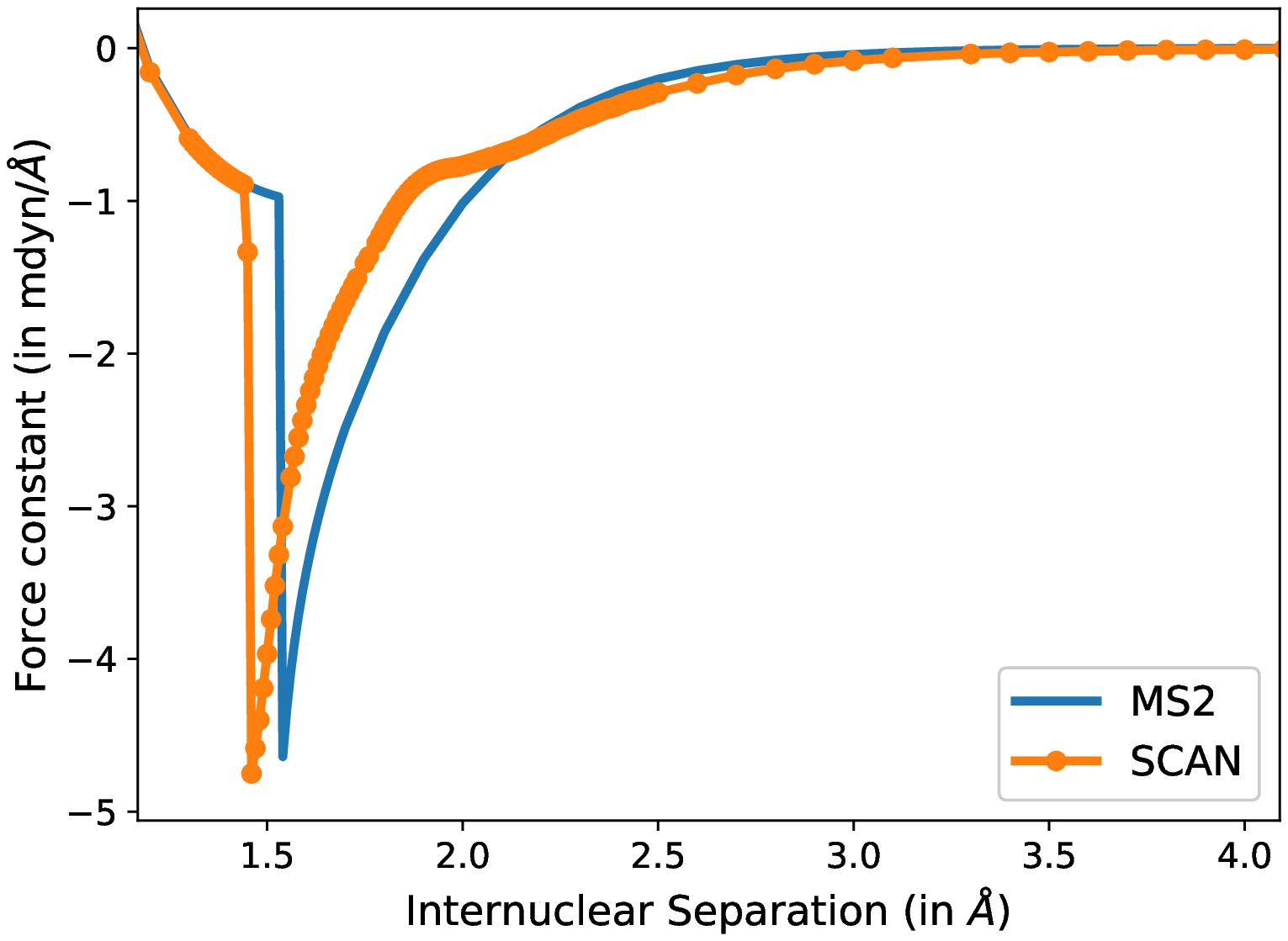}
    		\end{minipage}
    		\begin{minipage}{0.48\textwidth}
    			\includegraphics[width=\linewidth]{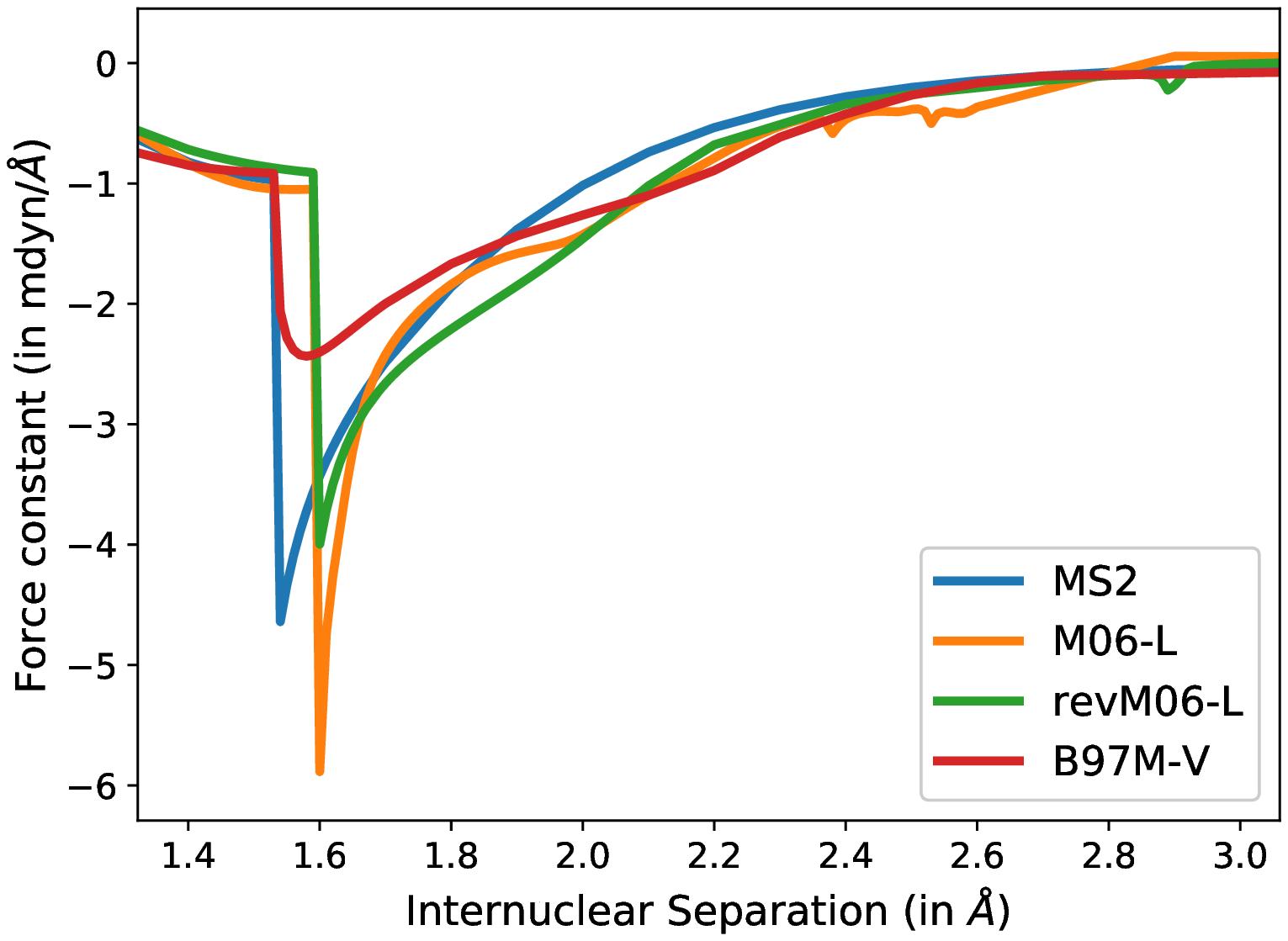}
    		\end{minipage}
    		\caption{Force constant for bond stretching predicted by SCAN (left) and some empirical mGGAs (right), with an acceptable functional from the same rung of Jacob's ladder as a reference.}
    		\label{fig:fconstinacc2}
    	\end{figure}

In contrast, SCAN and the empirical mGGAs M06-L, revM06-L and B97M-V appear to possess a normal discontinuity at their CF points, though ripple like features are present further down the dissociation curves. None of them possess any evident points of catastrophic failure, though their behavior is clearly suboptimal relative to behavior predicted by MS2 (which is qualitatively consistent with exact quantum mechanics, aside from the CF point discontinuity). 

\subsection{Relation to lowest eigenvalue of orbital rotation Hessian}

Eqn \ref{eqnpol} shows a connection between second order properties (like force constants and polarizabilities) and the inverse of the orbital rotation Hessian matrix. The smallest eigenvalue of the Hessian corresponds to the largest eigenvalue in the Hessian inverse, and consequently could have disproportionate impact on the second order property predictions. It is therefore interesting to study the behavior of the smallest Hessian eigenvalue over the dissociation curve, and compare/contrast with the observations in the preceding sections. 
	\begin{figure}[htb!]
			\includegraphics[width=0.5\linewidth]{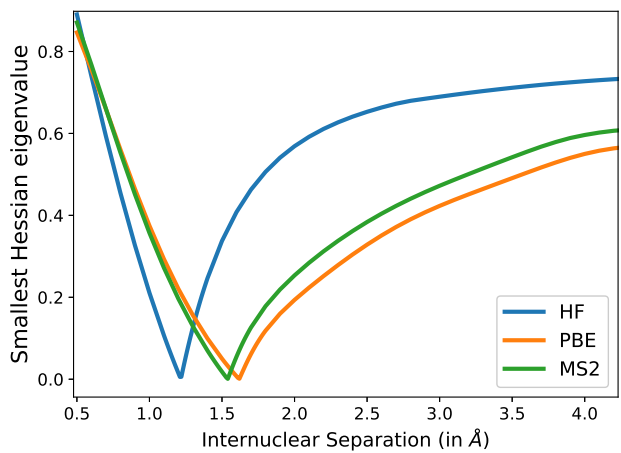}
		\caption{Smallest Hessian eigenvalue for HF, PBE and MS2.}
		\label{fig:hessianacc}
	\end{figure}

Fig \ref{fig:hessianacc} shows the smallest Hessian eigenvalue for the three qualitatively acceptable methods HF, PBE and MS2. The behavior is largely as expected, with the eigenvalue decreasing with increasing bond stretch until the CF point, where it becomes zero to permit barrierless transition to the spin polarized state. The effects of the transition manifest themselves in the form of a kink as the two segments of the curve belong to qualitatively different (restricted vs spin polarized) solutions. The eigenvalue increases in magnitude after the CF point due to increasing stability of the spin--polarized solution, ultimately asymptoting to the atomic limit. 

\begin{figure}[htb!]
		\begin{minipage}{0.48\textwidth}
			\includegraphics[width=\linewidth]{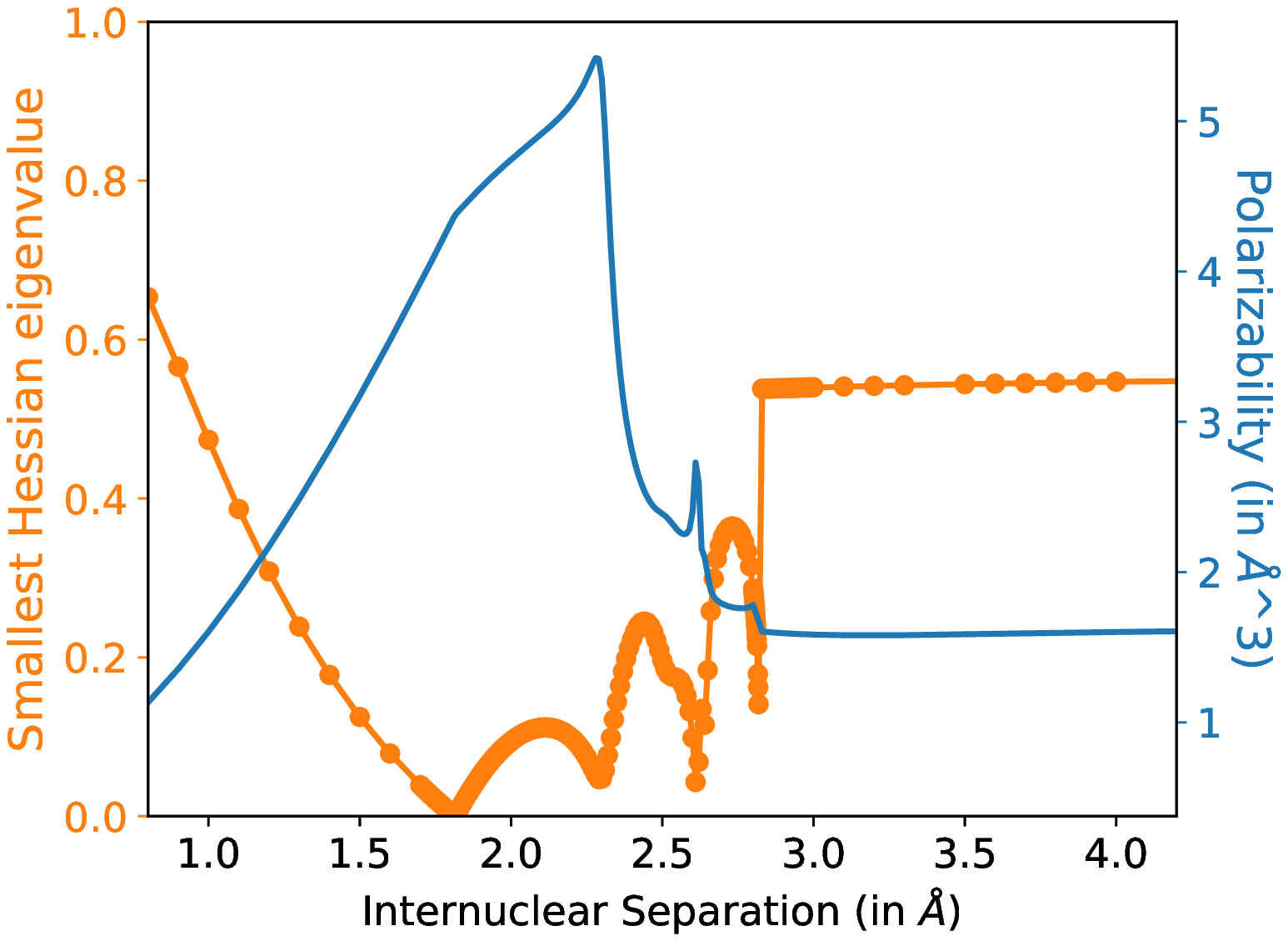}
		\end{minipage}
		\begin{minipage}{0.48\textwidth}
			\includegraphics[width=\linewidth]{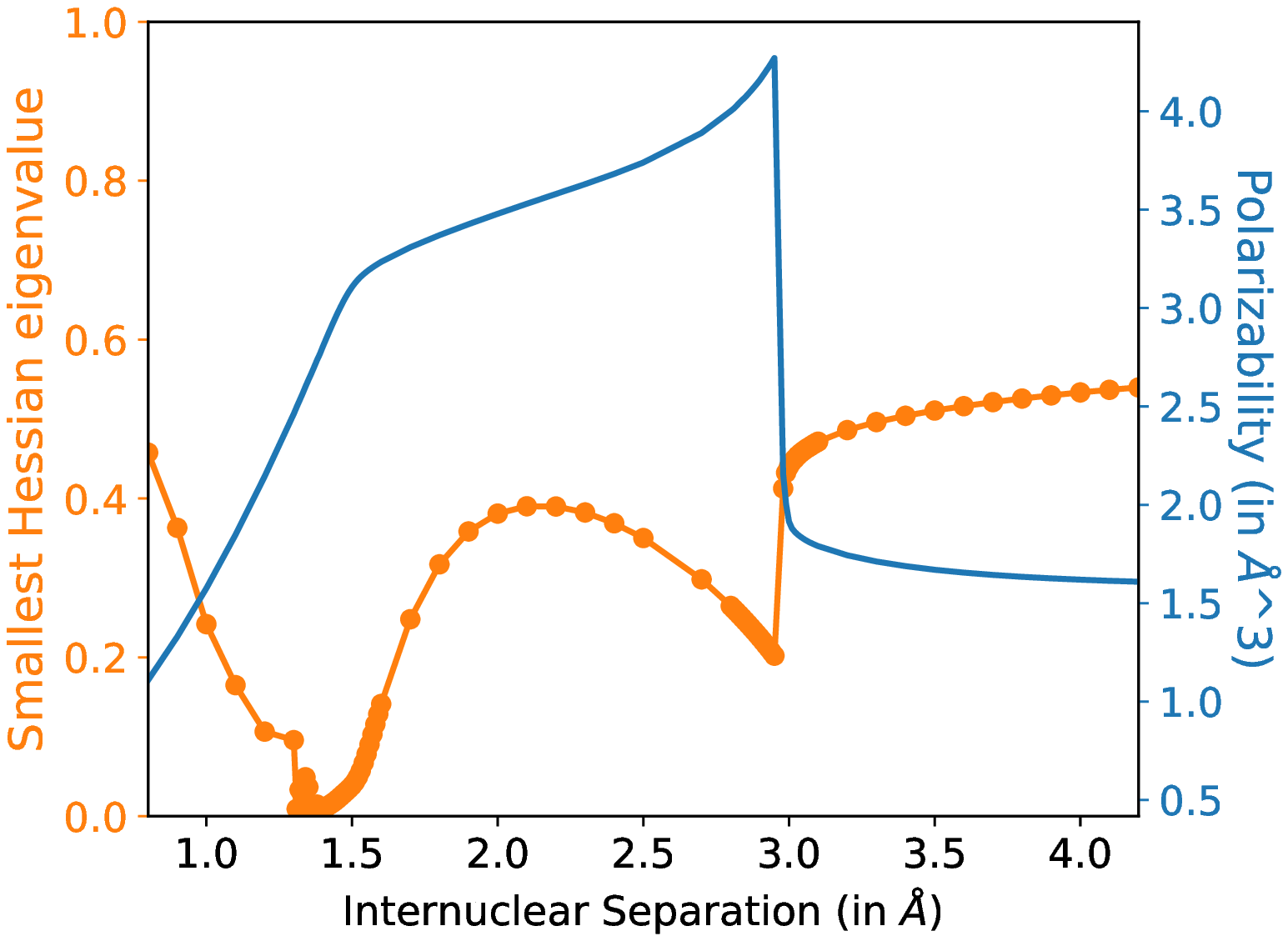}
		\end{minipage}
		\caption{Smallest orbital rotation eigenvalues of B97-D (left) and TPSS (right), alongside polarizability predictions.}
		\label{fig:hessb97d}
	\end{figure}

\begin{figure}[htb!]
		\begin{minipage}{0.48\textwidth}
			\includegraphics[width=\linewidth]{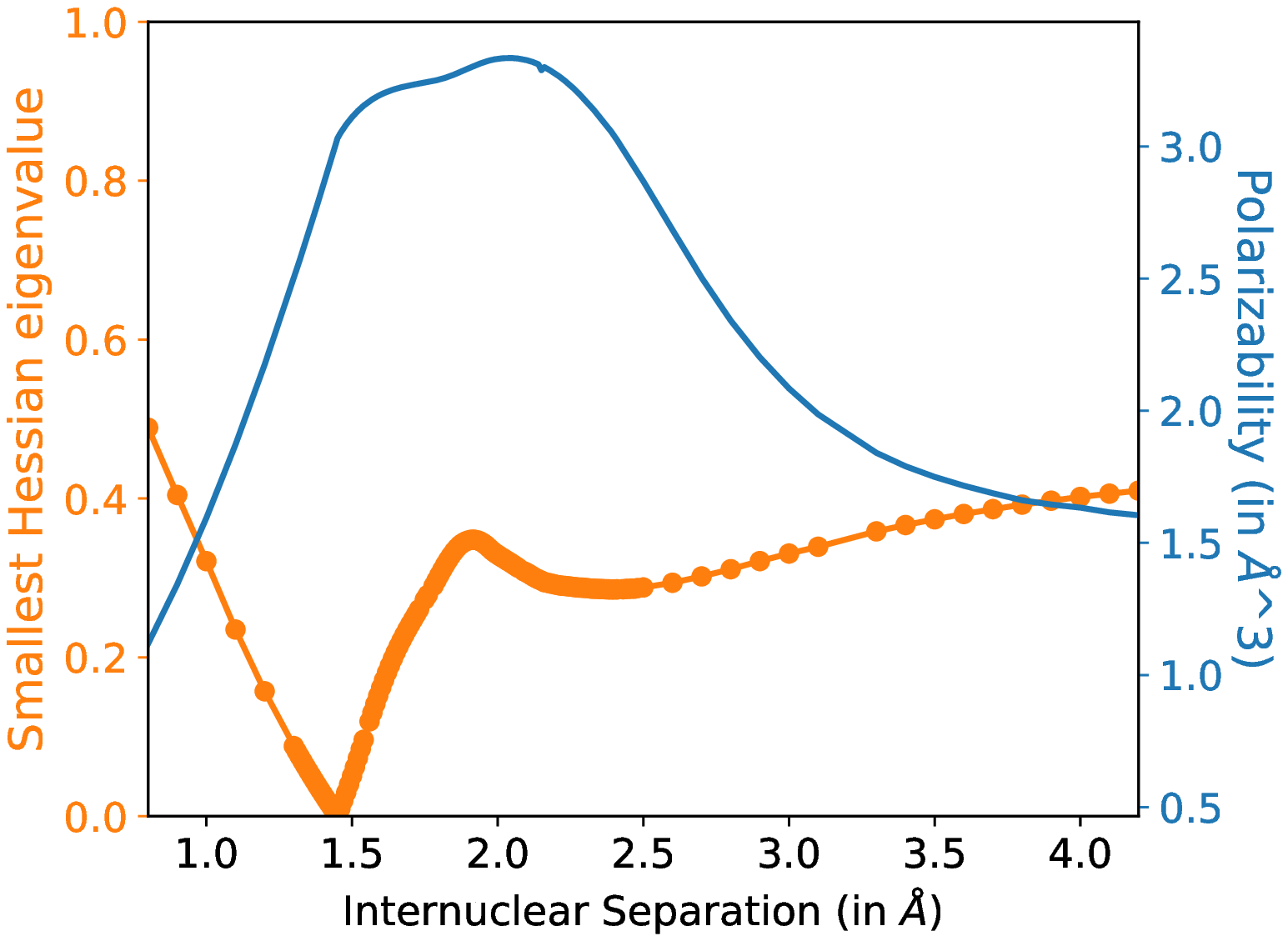}
		\end{minipage}
		\begin{minipage}{0.48\textwidth}
			\includegraphics[width=\linewidth]{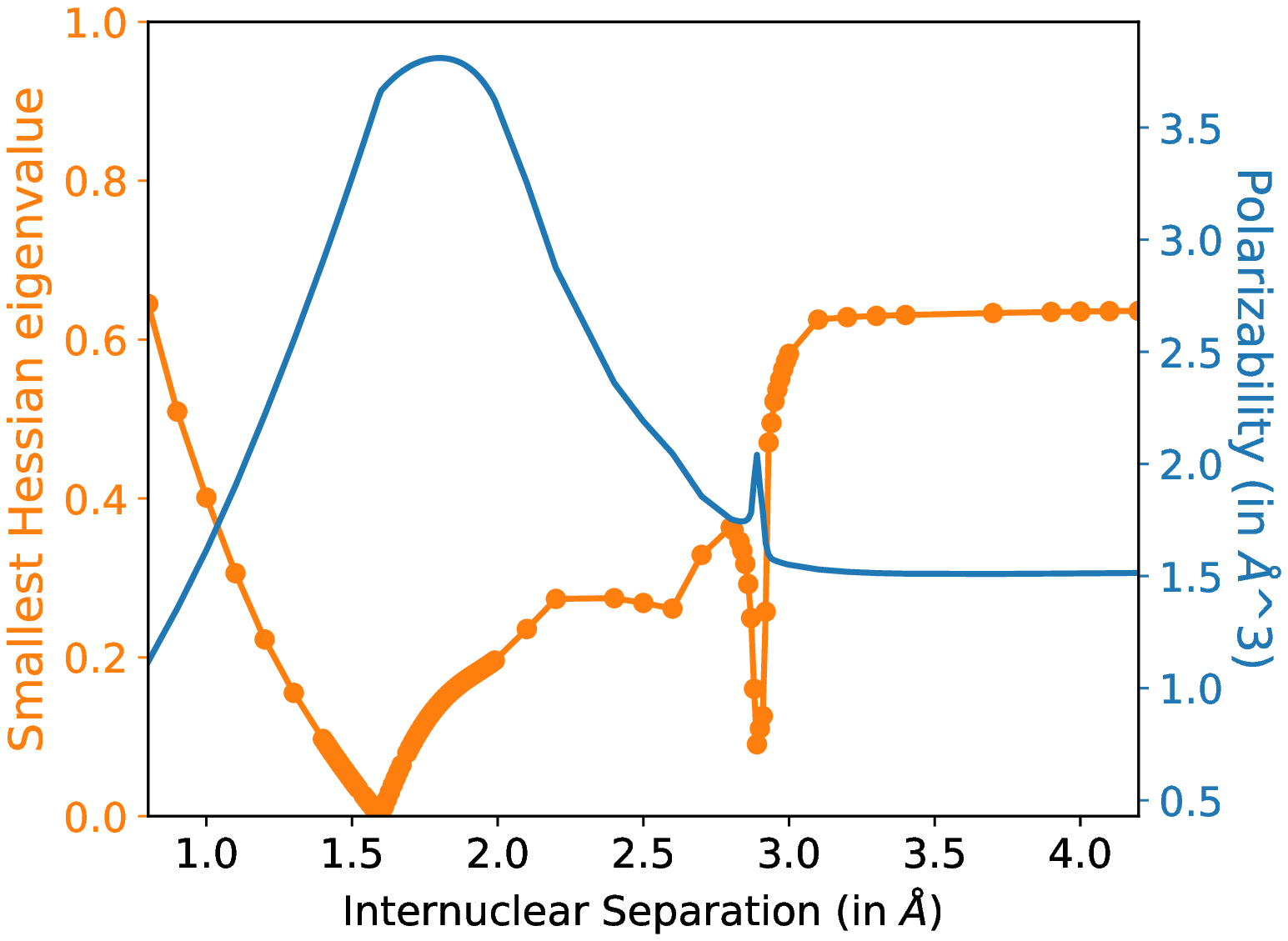}
		\end{minipage}
		\caption{Smallest orbital rotation eigenvalues of SCAN (left) and revM06-L (right), alongside polarizability predictions.}
		\label{fig:hessSCAN}
	\end{figure}

The problematic functionals, on the other hand, yield less sensible behavior, as can be seen from Figs 
\ref{fig:hessb97d} and \ref{fig:hessSCAN}. Critically however, the unphysical features in the polarizability (and force constant plots) are exactly reflected by the smallest Hessian eigenvalue, indicating that those features emerge from the orbital rotation Hessian inverse term in Eqn \ref{eqnpol}, as opposed to the other terms. Indeed, several features that were somewhat muted in the polarizability plots (like the small discontinuity around $r=3$ {\AA} in B97-D ) are considerably magnified by the eigenvalue plot, and are easier to see. In particular, the eigenvalue plot in Fig \ref{fig:hessSCAN} clearly reveals how the double peaked structure for the SCAN functional arises from non-monotonicity of the smallest Hessian eigenvalue. Issues with revM06-L that were not immediately apparent from the polarizability plot are also exposed by the zig-zag behavior of the the eigenvalue. Finally, we note that the TPSS eigenvalue appears to also go to zero around the CF point, and does not really show unusual behavior immediately beyond it. However, the values appear to be somewhat oscillatory at bond lengths just shorter than the CF point, but this is not really reflected by the observables like the force constant. The origin of the unusual CF point behavior of TPSS therefore remains unresolved with this analysis. Nonetheless, the smallest orbital rotation Hessian eigenvalue proves to be an extremely useful metric in identifying unphysical features in second derivative properties, since it serves as a lodestar by greatly magnifying the errors present in the observables.

\color{black}
\subsection{Spin localization}

	\begin{figure}[htb!]
		\begin{minipage}{0.48\textwidth}
			\includegraphics[width=\linewidth]{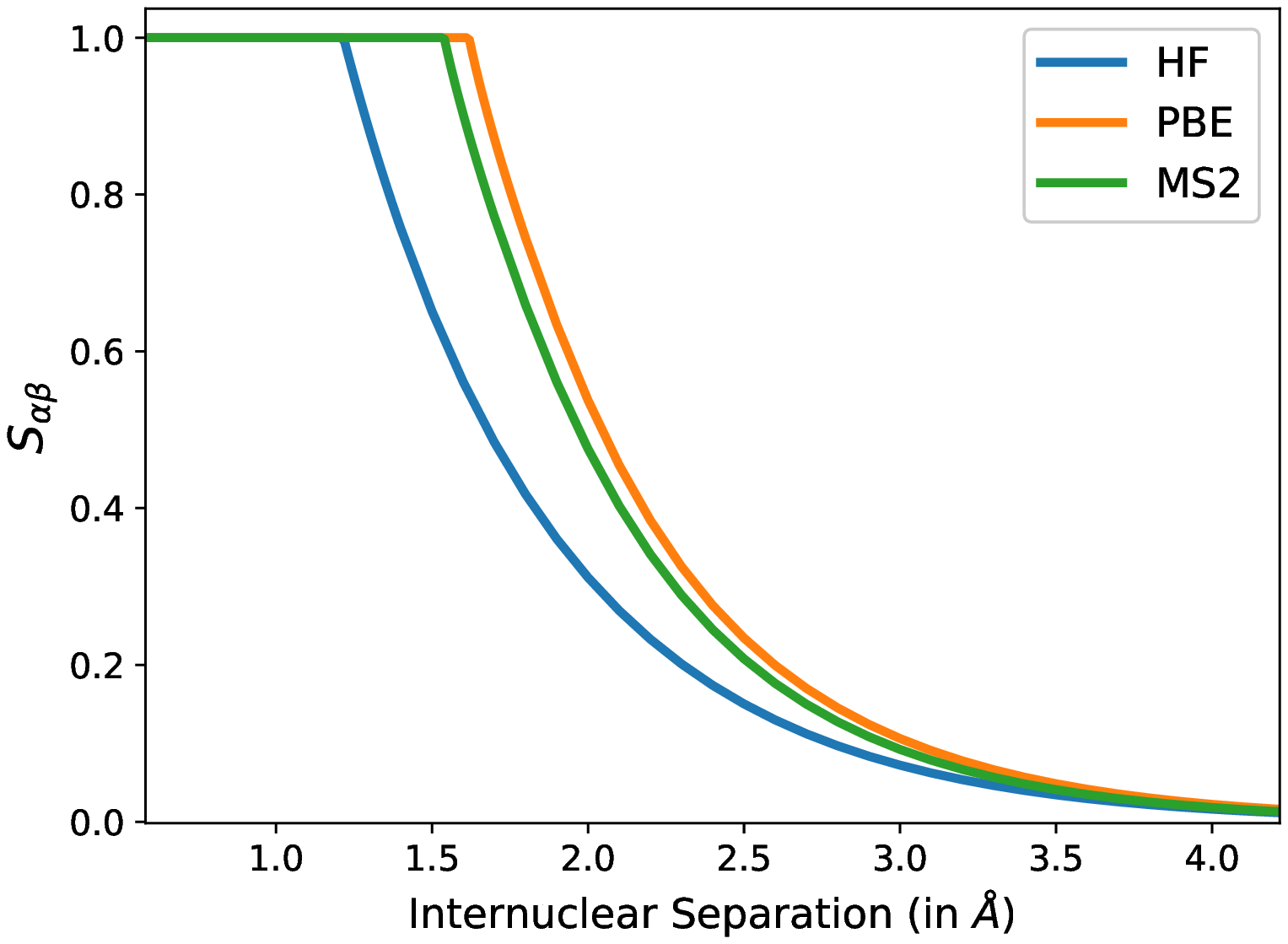}
		\end{minipage}
		\begin{minipage}{0.48\textwidth}
			\includegraphics[width=\linewidth]{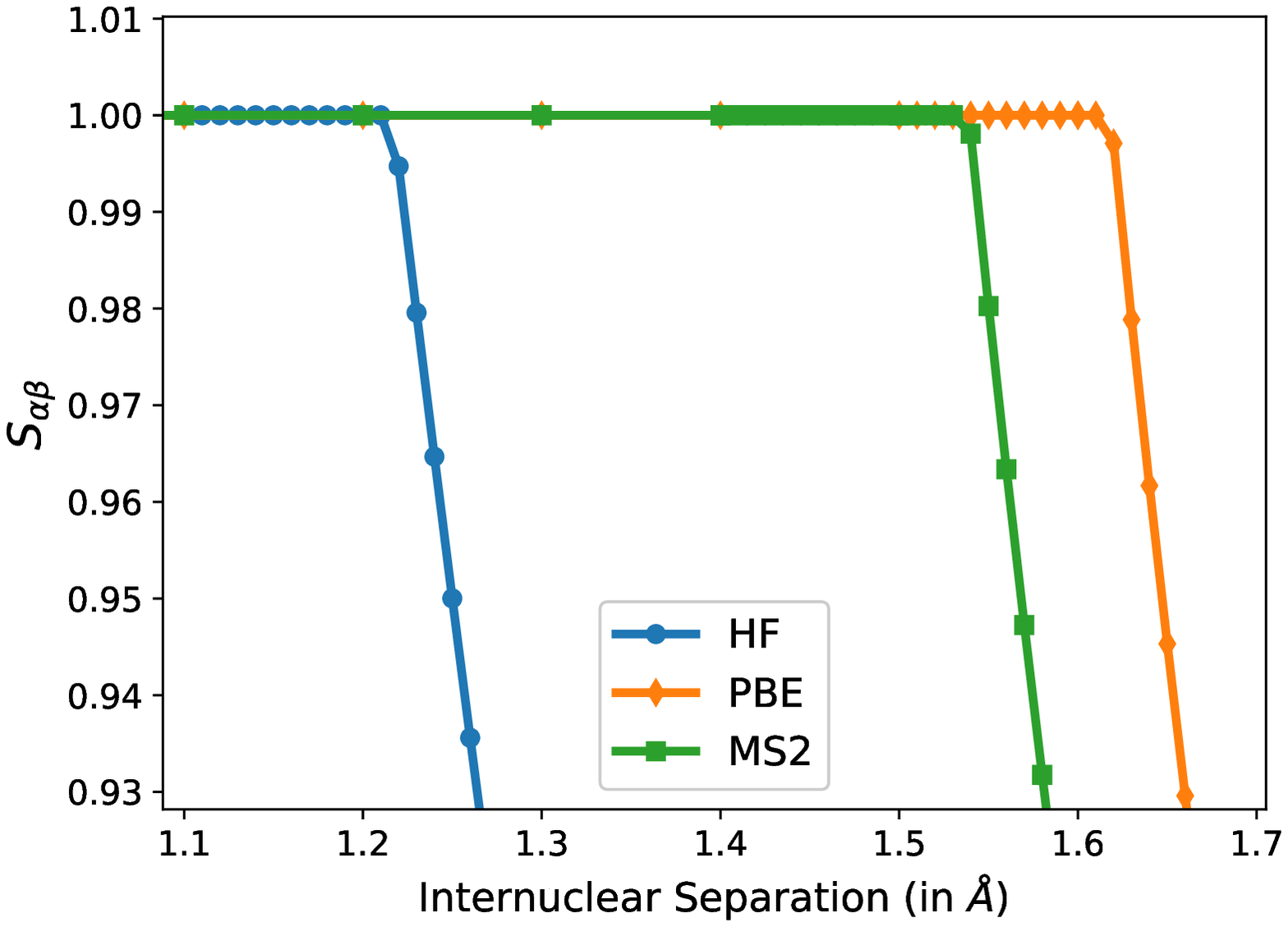}
		\end{minipage}
		\caption{Overlap $S_{\alpha\beta}$ predicted by three `qualitatively acceptable' methods (HF, PBE and MS2) for stretched H$_2$, highlighting a constant plateau till the CF point, followed by exponential decay (left). The right panel shows the behavior close to the CF point, highlighting a clear derivative discontinuity for all three methods.}
		\label{fig:overlapacc}
	\end{figure}

We have hereto conjectured that the unphysical features of the polarizability and force constant plots stem from improper spin density localization into atoms. It would therefore be useful to actually compare the unphysical regions of the plots with a metric for spin density localization. Two such scalar metrics are immediately apparent-the $\langle \hat{S}^2 \rangle$ for the KS determinant, and the overlap between the up and down spin densities. The former however is a problematic metric since none of the observables corresponding to the KS determinant need have any physical meaning aside from the density (which is constrained to be the exact density). The latter on the other hand is quite unambiguous in the case of H$_2$ as the up and down spin densities should have \textit{zero} overlap in the infinite separation limit (for a single determinant theory), while having perfect overlap till the CF point. Furthermore, the spin densities $\rho_{\alpha}(\vec{r})=\abs{\phi_{\alpha}(\vec{r})}^2$ and $\rho_{\beta}(\vec{r})=\abs{\phi_{\beta}(\vec{r})}^2$ overlap in a manner that is easily calculated from orbital overlap. Specifically, the spatial components of the KS occupied orbitals $\phi_{\alpha}(\vec{r})$ and $\phi_{\beta}(\vec{r})$ are nodeless and real for H$_2$ (on account of being the lowest energy orbitals overall), indicating that the orbital overlap:
\begin{align}
S_{\alpha\beta}&=\abs{\displaystyle\int \phi_{\alpha}(\vec{r})\phi_{\beta}(\vec{r})d\vec{r}}=\displaystyle\int \sqrt{\rho_{\alpha}(\vec{r})\rho_{\beta}(\vec{r})}d\vec{r}
\end{align}
is a convenient measure of the density overlap. It is therefore reasonable to anticipate that $S_{\alpha\beta}$ will be $1$ till the CF point (as the spatial orbitals for the two spins will be identical), and would decay exponentially in the asymptotic limit, solely on account of the overlap between the decaying tails of the atomic orbitals. Indeed, our `qualitatively accurate' methods (HF, PBE and MS2) yield precisely this behavior, as can be seen from  Fig \ref{fig:overlapacc}.

	\begin{figure}[htb!]
		\begin{minipage}{0.48\textwidth}
			\includegraphics[width=\linewidth]{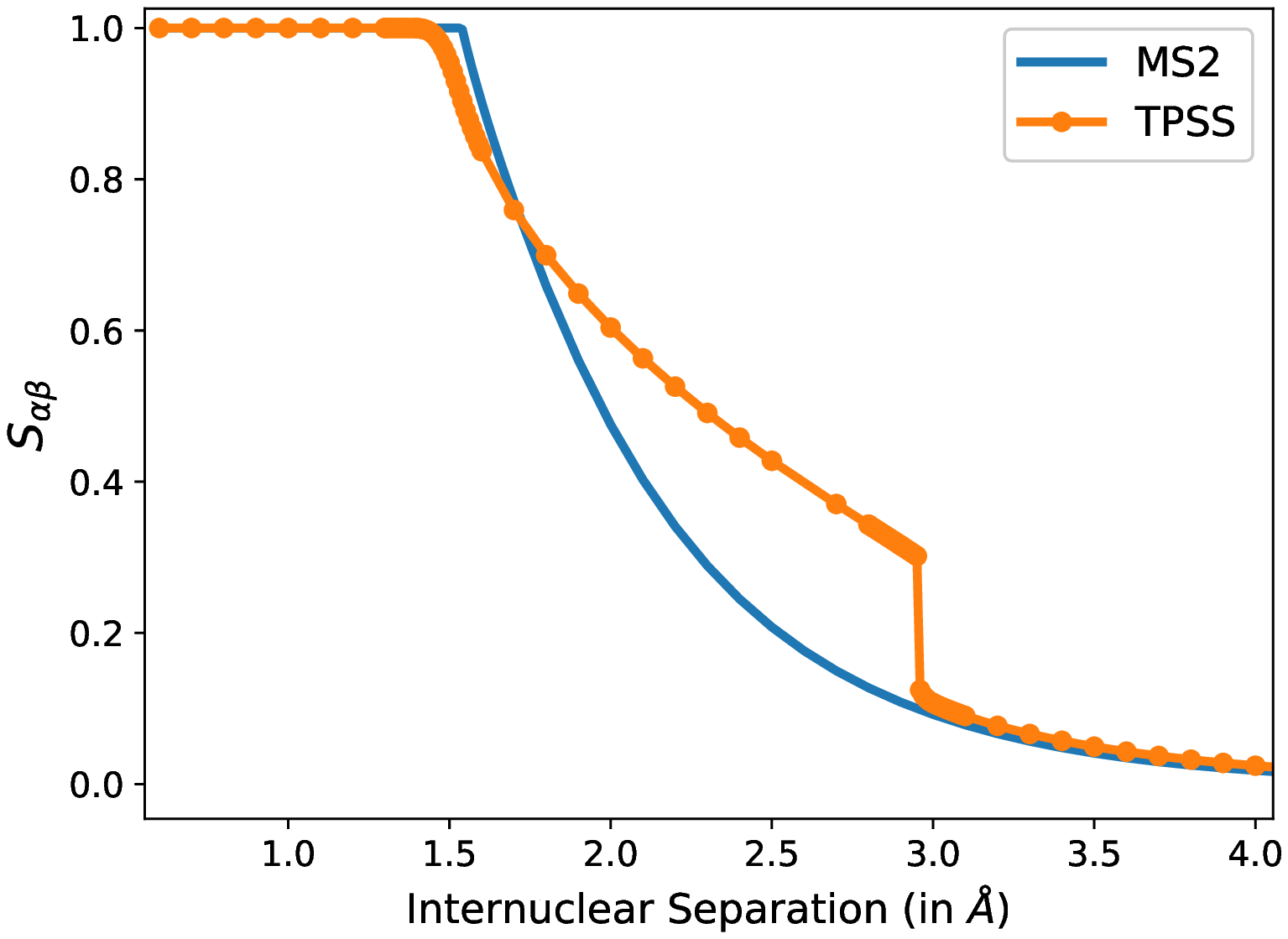}
		\end{minipage}
		\begin{minipage}{0.48\textwidth}
			\includegraphics[width=\linewidth]{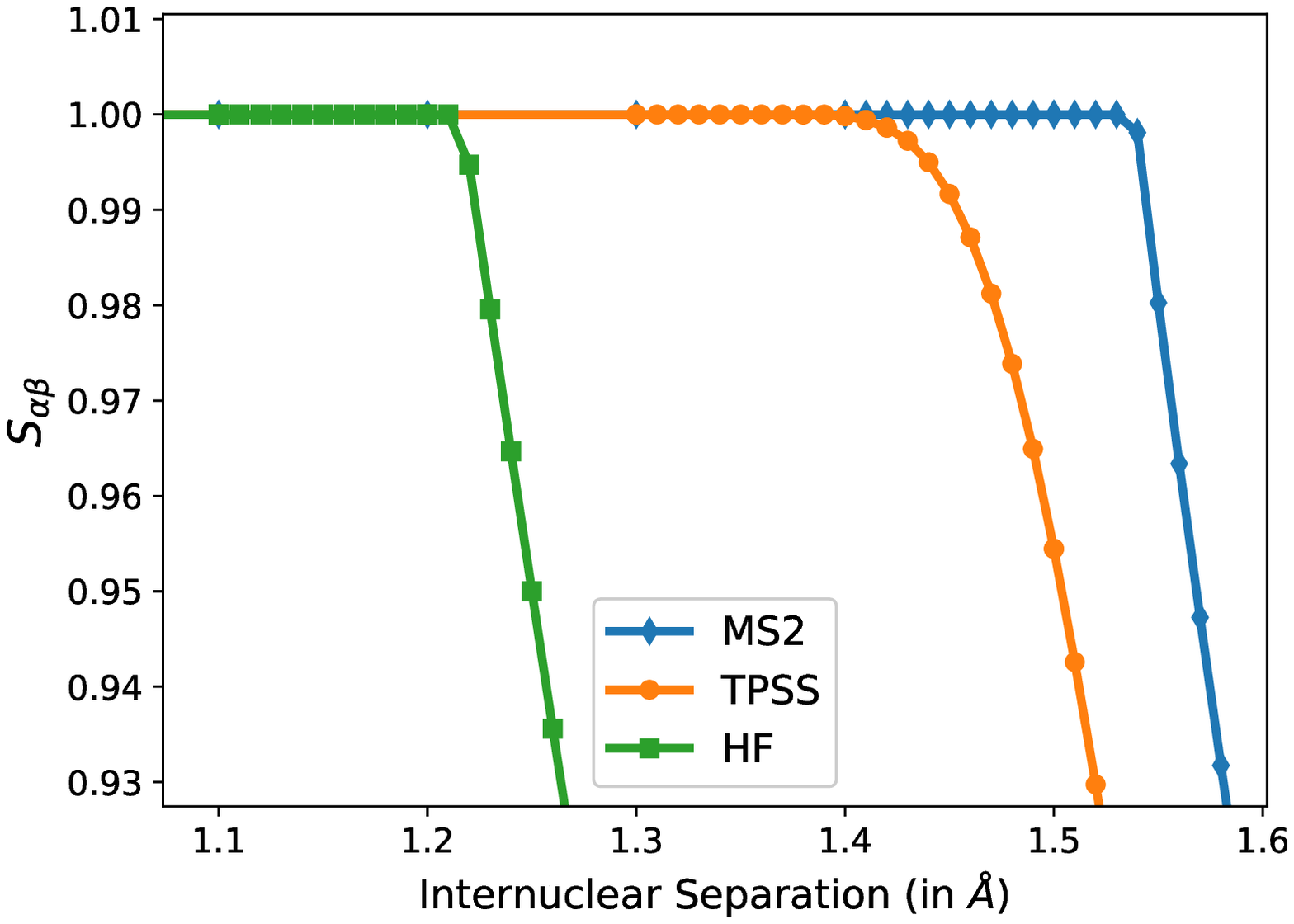}
		\end{minipage}
		\caption{Overlap $S_{\alpha\beta}$ generated by TPSS for stretched H$_2$, with some qualitatively accurate method(s) as reference(s). The left panel shows a clear discontinuity around $r=3$ {\AA}, which corresponds to discontinuities in the polarzability and force constant seen earlier. The right panel highlights lack of a derivative discontinuity at the CF point, unlike MS2 and HF. This is our strongest evidence showing that TPSS lacks a normal CF point.}
		\label{fig:overlaptpss}
	\end{figure}

We observe much more exotic behavior for some of the `problematic functionals' encountered earlier. TPSS (Fig \ref{fig:overlaptpss}) in particular shows several interesting features, including a discontinuity at 3 {\AA} indicating a discontinuous change in spin density. There is also a lack of an expected derivative discontinuity at the CF point. This seems to suggest that TPSS starts spin polarizing too slowly at the CF point, resulting in a partially spin polarized state till $r=3$ {\AA}. This incomplete spin polarization consequently preserves a partial covalent bond between the atoms, leading to large polarizabilities of the form seen in Fig \ref{fig:polarinacc}. The slow spin polarization also perhaps leads to $\dfrac{\partial ^2E}{\partial\theta \partial x}=0$ at the CF point along the zero eigenvalue Hessian mode, voiding the singular term in the Eqn \ref{eqnpol} and thereby leading to a continuous force constant (unlike all other methods). It further appears that the asymptotic limit of the $S_{\alpha\beta}$ segment upto 3 {\AA} is $\approx 0.14$ as opposed to $0$ (from fitting to a functional form of $A+Be^{-cx}$), indicating that the state in question corresponds to an entirely different, partially spin polarized UKS state  as opposed to the fully polarized state that is the standard solution to the UKS equations beyond the CF point.  However, a discontinuous transition around $r= 3$ {\AA} to the fully spin-polarized, bond free state occurs, taking TPSS to the right asymptotic limit of independent atoms. This therefore suggests that there are two potential UKS states predicted by TPSS for stretched H$_2$, one that is fully spin polarized and one that is partially so, with the latter being energetically preferred at intermediate bond stretch levels. 

	\begin{figure}[htb!]
		\begin{minipage}{0.48\textwidth}
			\includegraphics[width=\linewidth]{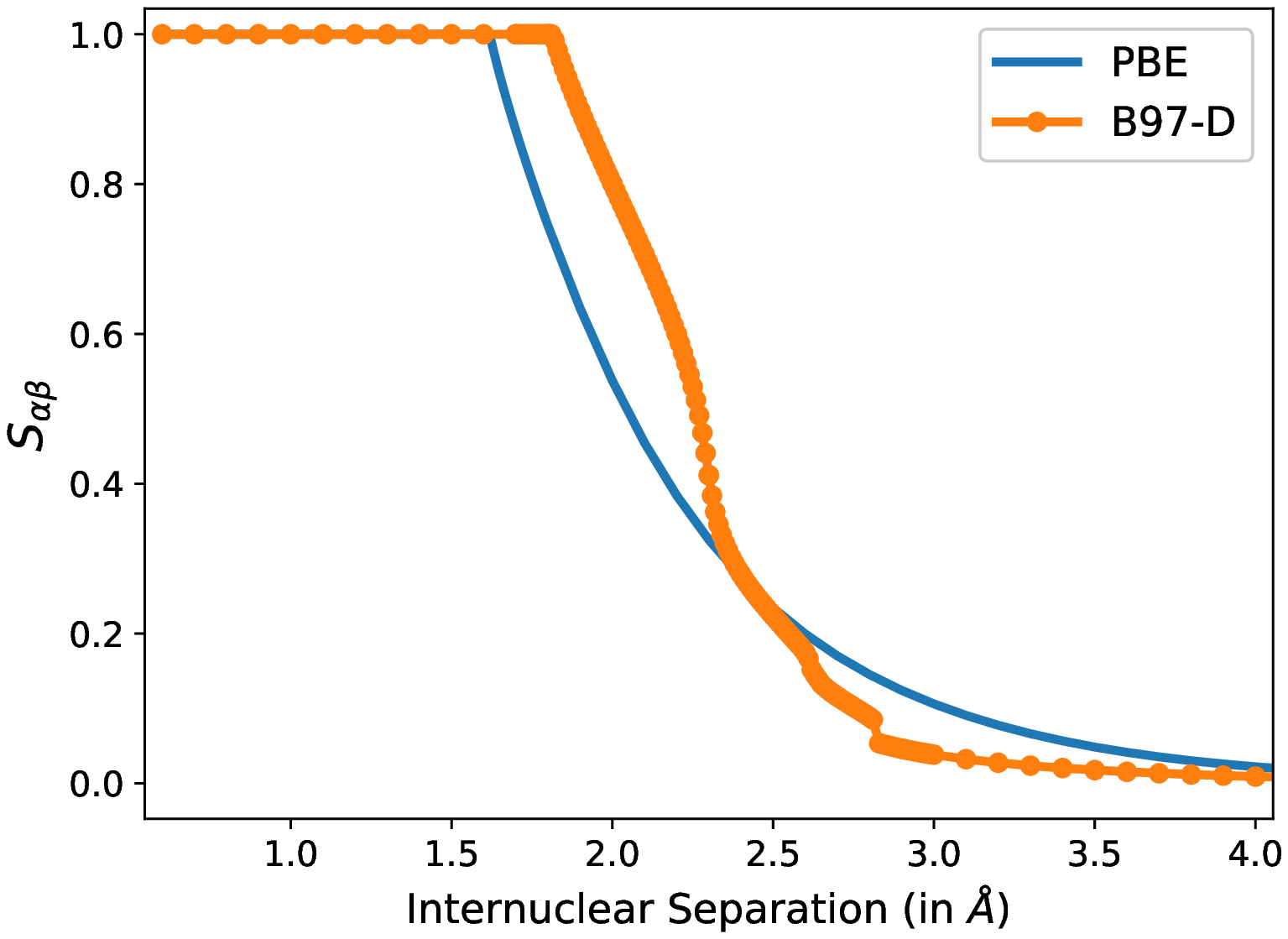}
		\end{minipage}
		\begin{minipage}{0.48\textwidth}
			\includegraphics[width=\linewidth]{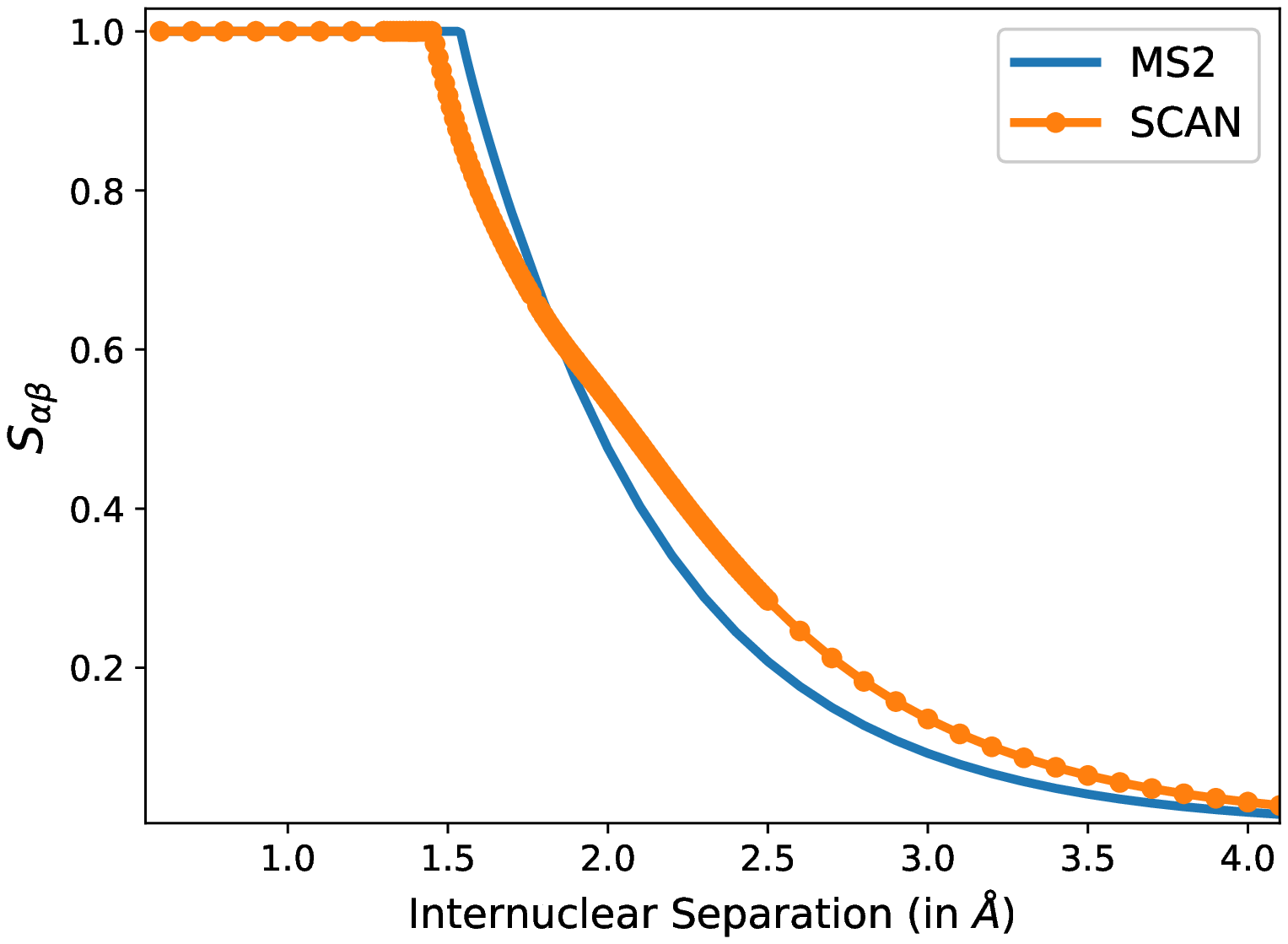}
		\end{minipage}
		\caption{Overlap $S_{\alpha\beta}$ generated by B97-D (left) and SCAN (right) for stretched H$_2$, with some qualitatively accurate method(s) as reference(s).}
		\label{fig:overlapothers}
	\end{figure}
	
$S_{\alpha\beta}$ for B97-D similarly shows some unexpected features like changes in  the sign of curvature at different points in the curve (Fig \ref{fig:overlapothers}, left panel), a very rapid (and non-exponential) decay around 2.3 {\AA} (which corresponds to the rapid drop in polarizability and the second peak in the force constant), some smaller staircase like features and a very small discontinuity around 2.8 {\AA}. The region of rapid drop in overlap seems to suggest a transition from a partially polarized state with a fractional bond to a fully polarized one, although the continuous nature of the change seems to suggest that it is a single state that smoothly changes its character, unlike the case of TPSS with a discontinuous transition. On the other hand, SCAN  has more reasonable behavior (Fig \ref{fig:overlapothers}, right panel), though there seems to be signs of a subtle change in curvature sign around $r= 2$ {\AA}, which matches nicely with the not so subtle behavior polarizability and force constants in that neighborhood, as shown earlier. Finally, the other, empirically fitted functionals like M06-L and B97M-V do not appear to show any strongly objectionable issues in their overlaps, though some small features are present. M06-L has a small ripple (i.e. change in curvature sign) around 2.4 {\AA}, corresponding to a sudden spike in the polarizability prediction and revM06-L has a small kink around 2.8 {\AA}, again corresponding to a spurious peak in the polarizability.  These subtle features are however nowhere as dramatic as the behavior seen for TPSS and B97-D, and on the whole, their performance is mostly satisfactory.

\subsection{Density-corrected polarizabilities}
The preceding section strongly suggests that unphysical predictions made by the functionals stem from inaccuracies in the underlying density (or more specifically, from the limited extent of spin polarization). It therefore seems instructive to consider the behavior of these functionals when supplied with a qualitatively correct density. We achieve this by calculating static polarizabilities (as predicted from the second order response of the energy to an applied electric field) from HF orbitals, using various functionals. This is essentially the density corrected DFT (DC-DFT) protocol proposed by Kim et al.\cite{kim2013understanding,kim2014ions}.

	\begin{figure}[htb!]
		\begin{minipage}{0.48\textwidth}
		\includegraphics[width=\linewidth]{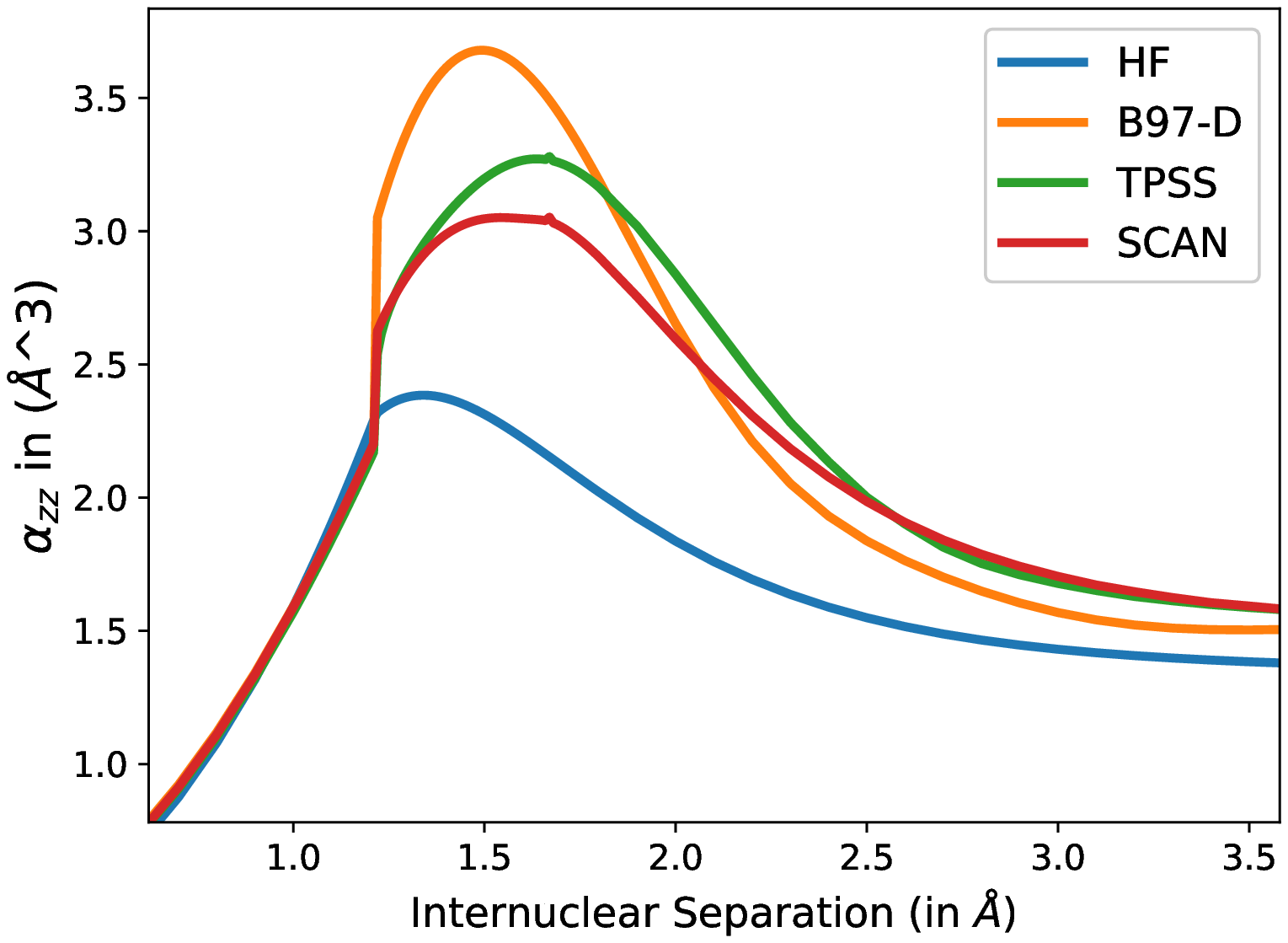}
		\end{minipage}
		\begin{minipage}{0.48\textwidth}
		\includegraphics[width=\linewidth]{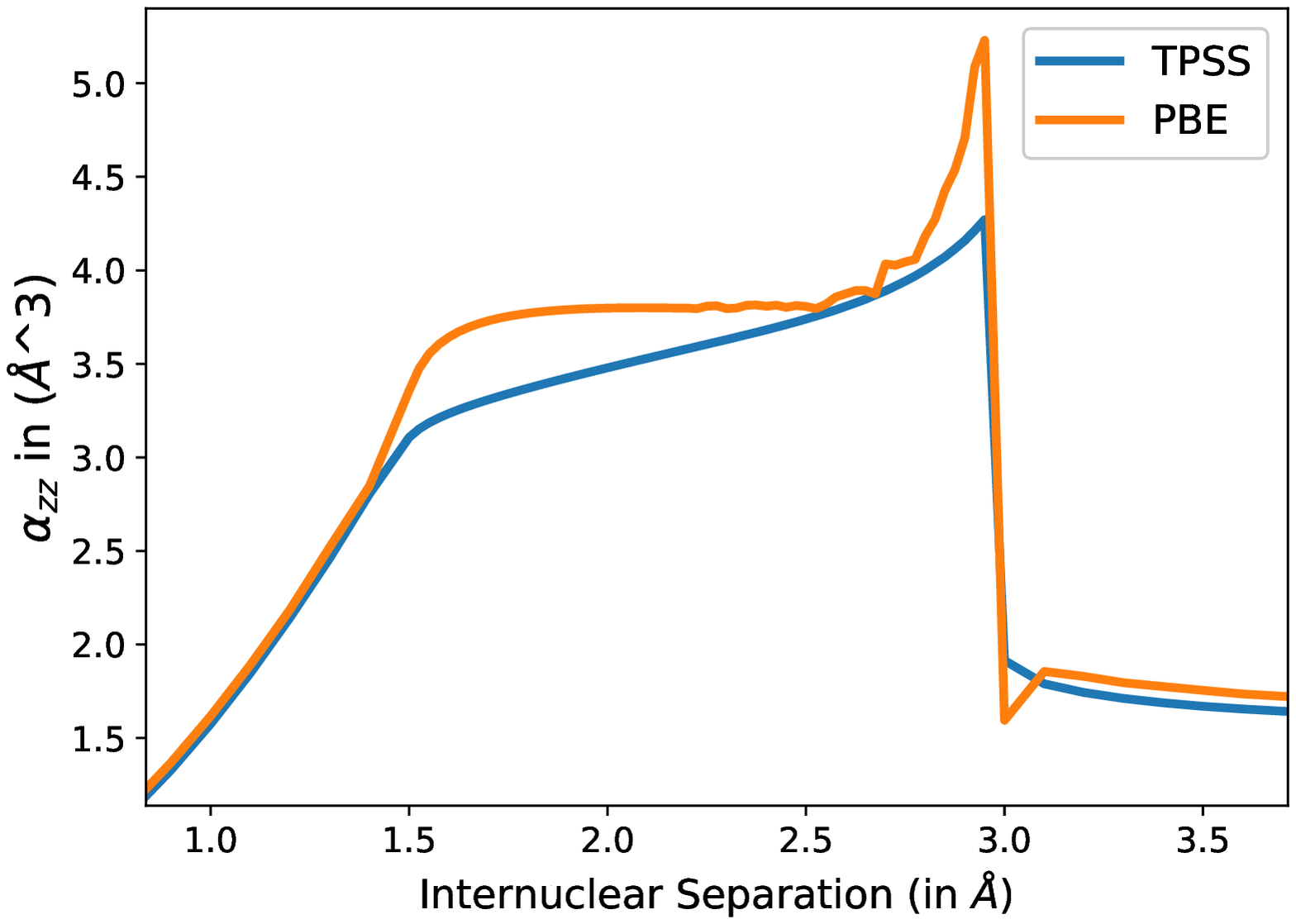}
		\end{minipage}
		\caption{Static polarizability parallel to the bond axis generated from HF orbitals (left) and TPSS orbitals (right).}
		\label{fig:notSCF}
	\end{figure}

Fig \ref{fig:notSCF} (left panel) shows that the unphysical features in the polarizability predictions of B97-D, TPSS and SCAN disappear once the UHF orbitals are used  instead of the self-consistent ones. There is however a clear discontinuity at the HF CF point, but this was largely expected due to the presence of terms that did not contribute to Eqn \ref{eqnpol} on account of self-consistency enforcing the Hellman-Feynman theorem. The behavior otherwise is quite reasonable, and supports the hypothesis that the polarizability errors stem from inaccuracies in the self-consistent density. Conversely, using the TPSS density to predict polarizability causes even the usually well behaved PBE functional to break down (Fig \ref{fig:notSCF}, right panel) and provide a similar shape as TPSS with self-consistent orbitals. We note that there is no CF point discontinuity when the TPSS density is used (though there is a barely perceptible kink) because TPSS does not have a normal CF point with a derivative discontinuity in $S_{\alpha\beta}$. We also observed much worse breakdown in polarizability predictions (including unphysical negative values!) when HF was combined with TPSS orbitals, or when PBE/HF was combined with B97-D orbitals, indicating that those densities were not suitable for generally well-behaved methods.

\subsection{Barriers to H atom association}
There are no local maxima in the energy of a stretched H$_2$ molecule, as predicted by exact quantum mechanics. Such a maximum would in fact not make physical sense, as it would indicate that there is a barrier to two H atoms associating to form H$_2$ and so there would be regions where the forces would push two H atoms further apart instead of closer together.

\begin{figure}[htb!]
	\includegraphics[width=0.5\linewidth]{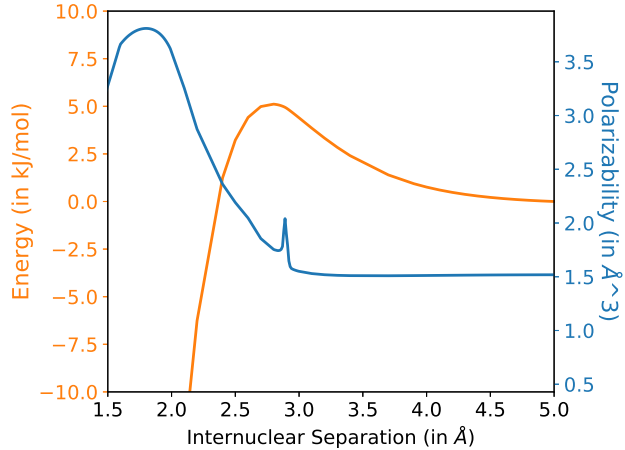}
	\caption{Energy of stretched H$_2$ relative to the dissociation limit, predicted by revM06-L, alongside static polarizability. The unphysical barrier and the strange polarizability spike occur in the same region of the stretch. It is possible that the polarizability spike results from a dipole allowed, low lying excited state, which perhaps mixes into the UKS determinant and spuriously elevates ground state energy.}
	\label{fig:barrier}
\end{figure}

While spurious barriers in dissociation curves are not unknown, they tend to arise mostly from non-variational optimization of parameters in methods like projected coupled cluster\cite{van2000benchmark} or xDH double hybrid functionals\cite{hait2018communication}. We however find that many density functionals appear to predict an unphysical barrier to association for two H atoms,\textit{ despite } self-consistent optimization of orbitals. So far we have noted the presence of a barrier in B97-D (11.4 kJ/mol), M06-L (13.5 kJ/mol), revM06-L (5.1 kJ/mol) and B97M-V (1.1 kJ/mol) local functionals. HF, PBE, TPSS, SCAN and MS2 remain blissfully barrier free, as do the hybrid methods TPSSh, SCAN0, PBE0, B3LYP and MS2h. A more extensive analysis of other popular and recent functionals (especially hybrid functionals) is presently underway, and it would be unsurprising if a fair number of those turn out to have spurious barriers as well.

We are not entirely sure if the barriers are necessarily connected to the polarizability catastrophes mentioned earlier. TPSS for instance lacks a barrier, despite having significant issues with polarizabilities. On the other hand, it is certainly possible that fractionally bonded states generated by bond stretches elevate the energy spuriously at intermediate bond stretches due to the presence of high energy ionic contributions. Both revM06-L and M06-L have the energy maximum in the neighborhood of their unphysical polarizability spikes (as can be seen for the former in \ref{fig:barrier}), supporting this view. However, the barriers do not disappear when the energy is calculated with HF orbitals instead of self-consistent ones (indeed, they marginally get bigger), indicating that is not just a failure of the self-consistent density. It must however be kept in mind that the use of non self-consistent densities elevates absolute energies, and is likely to have maximum impact at points where the self-consistent density is most different from the supplied density, which are the problematic points along the dissociation curve. So, it is perhaps expected that the barriers will persist even after density correction.  Overall, we are not quite sure of the reasons behind the origin of these barriers at present, but would recommend caution in applying the problematic functionals for ab-initio dynamics due to the potential of obtaining spurious forces and unphysical barriers.

\section{Discussion \& Conclusions}
KS-DFT methods are typically accurate and efficient for treating electronic structure in the absence of strong correlation effects. Prototypical examples include closed-shell molecules and their separated, spin-polarized, radical fragments. Yet in applications like \textit{ab initio} dynamics, such bonds may be made or broken, and density functionals may therefore have to cope with the strongly correlated spin-recoupling regime. The closed shell bonded regime and the spin-polarized broken bond regime are joined via the Coulson-Fisher (CF) point, which signals the onset of strong correlations that cause spin polarization for approximate single reference methods. High accuracy in the recoupling regime is not readily available today,but it is reasonable to expect that UKS models should smoothly join the CF point to the broken bond regime, similar to UHF. Such functionals can be considered to be well-behaved.  

In this work, we have examined the homolytic bond-breaking of the H-H bond as a simple prototype, using second derivative properties (polarizabilities and force constants) at stretched bond-lengths as sensitive probes of whether or not functionals are well behaved. We consequently find that the reasonable expectation that UKS models can describe the dissociation of H$_2$ (a nonpolar, singly bonded neutral molecule) with qualitative accuracy to be flawed. A number of popular (PBE, B3LYP) and recent (MS2) functionals can in fact describe properties with similar qualitative features as exact quantum mechanics, but their success is by no means indicative of the general UKS performance. A fair number of functionals (both empirically trained and non-empirically constrained) appear to yield unphysical predictions for static polarizability and force constant for bond stretching. The most egregious offenders are the empirically fitted B97-D GGA and the nonempirical TPSS meta GGA, which appear to predict a fractionally bonded state with incomplete spin localization and high polarizability at intermediate bond stretch levels, and subsequently undergo a dramatic transition to the atomic asymptotic limit with localized spins. Other functionals like SCAN and M06-L have fewer objectionable features, but their predictions do not appear physical at all times. All these unphysicalities appear to originate from the smallest eigenvalue of the orbital Hessian matrix, whose erratic behavior in turn appears to stem from incorrect levels of spin polarization in the self-consistent density. Consequently, the polarizability errors can be corrected via the use of physically correct, completely spin polarized densities such as those from UHF, indicating that the errors here are essentially `density driven'. 

We also find that a large number of functionals predict a spurious barrier impeding association of free H atoms to form H$_2$, although a clear connection between these barriers and the polarizability catastrophes cannot yet be clearly drawn. It is however quite clear that the presence of such barriers would be problematic for any potential application of these functionals in regimes involving bond breaking. In particular, \textit{ab initio} molecular dynamics simulations using these methods would suffer from incorrect forces and artificial barriers, potentially affecting the conclusions. We therefore recommend extensive investigation of the potential energy surface along the reaction coordinate before employing these functionals for dynamical studies, in order to ensure that unphysical effects are kept to a minimum. 

Overall, we have identified a new class of errors beyond delocalization error and static correlation that plagues a fair number of density functionals, preventing them from describing the dissociation of even simple species like H$_2$. These errors appear to mostly stem from inaccuracies in the underlying self-consistent density than an obvious defect of the functional form  itself. Further work is required to characterize this class of errors, in order to mitigate their occurrence in future functionals.

\section{Acknowledgements}
This research was supported by the Director, Office of Science, Office of Basic Energy Sciences, of the U.S. Department of Energy under Contract No. DE-AC02-05CH11231. D.H. was also funded by a Berkeley Fellowship and A.R. via the Berkeley Science Network.
	\bibliography{references}
\end{document}